\documentclass[preprint,12pt]{elsarticle}

\usepackage{amssymb,bm,amsmath}
\usepackage{svg}
\usepackage{booktabs}
\usepackage{multirow}
\usepackage{array} % Added for p{} column specifier
\usepackage{mwe}
\usepackage{graphbox} %loads graphicx package\usepackage{caption}
\usepackage{geometry}
\newcolumntype{M}[1]{>{\centering\arraybackslash}m{#1}}
\usepackage{subcaption}
\usepackage{comment}
\usepackage{algorithm}

\usepackage{array}
\usepackage{ragged2e}
\usepackage{mathtools}

\usepackage{algpseudocode}
\journal{Journal of Computational Physics}

\begin{document}

\begin{frontmatter}



%\title{Eulerian Interface Tracking with Triangle Edge Cuts}
%\title{A Particle Flow Map Method for Incompressible Flow}
%\title{An Incompressible Navier-Stokes Solver based on Long-range Particle Flow Maps}
\title{An Impulse-formed Navier-Stokes Solver based on Long-range Particle Flow Maps}


% \author[inst1]{Mengdi Wang\corref{cor1}}

% \cortext[cor1]{Corresponding author.}

% \fntext[emails]{E-mail addresses: mengdi.wang@gatech.edu (Mengdi Wang), matthew.d.cong@gmail.com (Matthew Cong), bo.zhu@gatech.edu (Bo Zhu).}

% \affiliation[inst1]{organization={School of Interactive Computing},%Department and Organization
%             addressline={Georgia Institute of Technology}, 
%             city={Atlanta},
%             postcode={30332}, 
%             state={Georgia},
%             country={USA}}

% \author[inst2]{Matthew Cong}

% \affiliation[inst2]{organization={NVIDIA Corporation},%Department and Organization
%             addressline={2788 San Tomas Expy}, 
%             city={Santa Clara},
%             postcode={95051}, 
%             state={California},
%             country={USA}}

% \author[inst1]{Bo Zhu}

% --- Authors ---
\author[inst1]{Zhiqi Li\fnref{cofirst}\corref{cor1}}
\author[inst1]{Duowen Chen\fnref{cofirst}\corref{cor1}}
\author[inst2]{Junwei Zhou\fnref{cofirst}}
\author[inst1]{Sinan Wang}
\author[inst1]{Yuchen Sun}
\author[inst1]{Bo Zhu}

% --- Notes ---
\fntext[cofirst]{These authors contributed equally as co-first authors.}

\cortext[cor1]{Corresponding author.}

\fntext[emails]{E-mail addresses:  zli3167@gatech.edu (Zhiqi Li),
dchen322@gatech.edu (Duowen Chen),
zjw330501@gmail.com (Junwei Zhou),
swang3081@gatech.edu (Sinan Wang),
yuchen.sun.eecs@gmail.com (Yuchen Sun),
bo.zhu@gatech.edu (Bo Zhu).}

% --- Affiliations ---
\affiliation[inst1]{organization={School of Interactive Computing},% Department and Organization
            addressline={Georgia Institute of Technology},
            city={Atlanta},
            postcode={30332},
            state={Georgia},
            country={USA}}

\affiliation[inst2]{organization={Computer Science and Engineering},% Department and Organization
            addressline={University of Michigan},
            city={Ann Arbor},
            postcode={48109},
            state={Michigan},
            country={USA}}

\begin{abstract}
We present a particle–grid characteristic-mapping framework that extends long-range characteristic mapping from inviscid flows to general Navier–Stokes dynamics with viscosity, body forces, and complex boundaries. Unlike traditional grid-based and vorticity-centered characteristic methods, our method is built on the observation that particle trajectories naturally provide the long-range flow map, enabling geometric quantities and their gradients to be transported in a direct and effective manner. We identify the impulse, the gauge variable of the velocity field, as the primary quantity mapped along characteristics while remaining compatible with standard velocity-based incompressible solvers. Using the 1-form representation of the impulse equation, we derive an integral formulation that decomposes the impulse evolution into a component transported geometrically along the particle flow map and a complementary component generated by viscosity and body forces evaluated through path integrals accumulated along particle trajectories. These components together yield a unified characteristic-mapping solver capable of handling incompressible Navier–Stokes flows with viscosity and body forces while maintaining the accuracy and geometric fidelity of characteristic transport.
\end{abstract}



\begin{keyword}
% Interface Tracking \sep Interface Representation \sep Volume of Fluid \sep Volume-conserving Advection Algorithm
Characteristic Mapping, Particle Flow Map, Impulse Variable,  Hybrid Particle-Grid methods, Navier-Stokes Equation
\end{keyword}

\end{frontmatter}


\section{Introduction}

A wide range of fluid dynamics problems involve the development of vortical structures and their interactions with other physical elements such as fluid interfaces, viscosity forces, and solid boundaries. Developing numerical methods that can accurately capture the generation of these vortices, preserve their coherent spatiotemporal structures, and model their interactions with surrounding physics has long been a central focus in computational fluid dynamics research. A multitude of numerical methods and computational programs have been developed to address the long-range evolution of vortical structures. As these algorithms are too extensive to provide a comprehensive survey here, we categorize them into a few main groups based on the key techniques employed: (1) high-order advection schemes (e.g., WENO \cite{liu1994weighted,hu1999weighted}, ENO \cite{harten1997uniformly,dumbser2017central}, BFECC \cite{dupont2003back,dupont2007back}, MacCormack \cite{maccormack2002effect}), which solve the advection equation with high-order accuracy; (2) high-resolution discretization methods (e.g., octree \cite{ando2020practical,laurmaa2016octree} or sparse grid \cite{setaluri2014spgrid}), along with many GPU-based parallelization strategies, which allocate computational resources to vorticity-intense regions to capture their evolution within practical computing budgets; (3) gauge methods \cite{sayeinterfacial,saye2017implicit,weinan2003gauge}, which solve gauge-form Navier-Stokes equations to decouple the projection step from velocity evolution in order to better preserve vortical features; (4) vortex methods (e.g., vortex-in-cell \cite{couet1981simulation,cottet2004advances}, vortex particles \cite{rossinelli2010gpu}), which solve the vorticity-form Navier-Stokes equations and evolve vorticity directly on Lagrangian particles; (5) gradient-augmented advection schemes such as Affine Particle-In-Cell (APIC) \cite{jiang2015affine,jiang2017angular} and Taylor Particle-In-Cell (Taylor PIC) \cite{nakamura2023taylor}, along with their variations for advecting other quantities such as level sets (e.g., \cite{kolomenskiy2016adaptive,kohno2013new,li2023garm}), which enhance advection accuracy by directly addressing the transport of high-order field quantities; and (6) geometric approaches, exemplified by characteristic mapping \cite{yin2021characteristic,yin2023characteristic,mercier2020characteristic}, which conserve quantities of interest by establishing long-range mappings from the original frame to the current frame.

Among these categories, geometric methods, featured by characteristic mapping methods (e.g., \cite{yin2021characteristic,yin2023characteristic,mercier2020characteristic,tessendorf2011characteristic}), have received extensive attention in fluid simulation across both scientific and visual computing communities. Characteristic mapping methods construct a mapping from the initial configuration to the current configuration by integrating the underlying velocity field along Lagrangian particle trajectories, thereby encoding the advection of quantities of interest through the deformation of the mapping. By explicitly tracking the evolution of the mapping function, these methods avoid numerical diffusion typically associated with grid-based advection schemes and enable long-range, high-fidelity transport of advected fields. A variety of characteristic mapping methods have been devised to solve incompressible flow problems based on different choices of mapping variables \cite{cortez1995impulse, yin2021characteristic, saye2016interfacial}, discretization strategies \cite{saye2016interfacial, cortez1996impulse}, and target physical couplings \cite{saye2016interfacial,saye2017implicit,chen2024solid,li2024ink}. A simple and classical example of characteristic mapping methods, which has been familiar to the fluid simulation community as a single-timestep mapping method, is the semi-Lagrangian method \cite{staniforth1991semi,strain1999semi,stam2023stable}. The semi-Lagrangian method approximates the characteristic mapping by tracing backward along the velocity field $\mathbf{u}$ to find the departure point $\mathbf{X}$, given by \(\mathbf{X} = \mathbf{x} - \int_{t-\Delta t}^{t} \mathbf{u}(\mathbf{x}(s), s) \, ds\), and interpolates the quantity from $\mathbf{X}$ to update the value at the current grid point. The time integration in computing the departure point is typically carried out using a numerical quadrature scheme, such as mid-point integration \cite{hammer1958midpoint}. While the semi-Lagrangian method enables unconditionally stable time integration by improving characteristic preservation, the entire mapping process occurs only within a single time step, and repeated interpolation from the grid to the back-traced virtual particles at each time step introduces additional numerical dissipation.

The characteristic mapping methods (e.g., \cite{yin2021characteristic,yin2023characteristic,mercier2020characteristic,tessendorf2011characteristic}) extend the concept of semi-Lagrangian advection from constructing a point-to-point mapping within a single timestep to establishing a long-range mapping over multiple timesteps. A typical approach to represent this mapping is by advecting the coordinates $\mathbf{X}$ discretized on grid nodes, so that each grid node at the current time $t$ stores its original coordinates at time $0$. The mapping for an arbitrary point $\mathbf{x}$ at time $t$ can then be obtained by interpolating the $\mathbf{X}$ values from the neighboring grid nodes. Quantities can be transported from time $0$ to $t$ directly by evaluating the interpolated values on the original frame at the query point $\mathbf{X}$. To enhance the mapping accuracy, the gradients of quantities are also mapped between the initial and current frames (e.g., see \cite{nave2010gradient,li2023garm,li2025edge,chen2025neural} %\bo{put gradient-augmented citations here Done}
). High-order interpolation schemes, such as fourth-order Hermite interpolation, are used to reconstruct the mapped variables based on both the mapped quantity and its gradient. Because the transported quantities and their gradients are read directly from the original frame at time $0$ rather than accumulated over multiple frames, no numerical dissipation is introduced during the process. As a result, the characteristic mapping approach is particularly suitable for solving problems that can be interpreted as geometric mapping problems, such as inviscid Euler flows (e.g., see \cite{yin2023characteristic,yin2021characteristic}). The same idea has also been used to solve level-set advection problems by preserving the interface features with the gradient-augmented map (e.g., \cite{nave2010gradient,li2023garm}).

Despite the notable success of characteristic mapping methods in solving geometric flow problems, the current literature still faces several challenges that hinder their use as a versatile numerical tool for solving general Navier–Stokes problems including all force terms and boundary conditions, as compared to traditional counterparts such as the grid-based advection-projection scheme \cite{staniforth1991semi,strain1999semi,stam2023stable,maccormack2002effect} or hybrid particle-grid methods \cite{jiang2015affine,jiang2017angular,nakamura2023taylor,couet1981simulation}. Three major challenges remain to be addressed for traditional characteristic mapping methods: 
(1) Most existing characteristic mapping methods maintain map information between frames by discretizing the map on two Eulerian grids, where the coordinates of the reference frame evolve with the velocity field in the current frame. Such a representation inevitably introduces numerical errors when interpolating the reference coordinates for arbitrary points within a grid cell, especially when the flow map exhibits non-trivial local stretching or compression that are not aligned with the grid lines. Therefore, devising new geometric representations that can accurately discretize the mapping by avoiding extensive grid-to-particle interpolation becomes an essential problem to address. 
(2) Current characteristic mapping methods lack mechanisms to handle quantities that cannot be incorporated into a purely mapping-based process, such as dissipative forces like viscosity, which require calculating a path integral along the particle trajectory between the initial and current frames. Because the current characteristic mapping method does not track the trajectory for each virtual particle, it is difficult to calculate and maintain the result of such integrals over a long period. Due to this missing component, the application scope of current characteristic mapping methods remains limited to solving fluid problems with ideal geometric mapping interpretations, such as vorticity transport in inviscid Euler flows \cite{yin2021characteristic,yin2023characteristic}, while general Navier–Stokes flows involving viscosity, body forces, and solid boundaries remain largely unexplored. 
(3) Taking the first two points into consideration (assuming solutions for each can be developed), it becomes essential to devise characteristic mapping methods as a two-step process: first identifying quantities that can be transported through the geometric mapping process, and then handling the remaining quantities through a path integral formulation. Consequently, it is critical to systematically review all gauge-form fluid equations and select the most appropriate formulation that fulfills these two requirements. In particular, what is needed to devise a successful characteristic-based Navier-Stokes solver is a gauge-form equation whose primary variable associated with the vorticity structure can be transported via a long-range flow map, while the remaining variables can be integrated in a unified manner along the long-range map discretization. Ideally, other core numerical techniques, such as field discretization, particle time integration, and Poisson-based pressure projection, should resemble those used in conventional incompressible solvers, enabling the method to leverage existing computational frameworks for solving complex flow problems.

In this paper, we propose a particle-based characteristic mapping method to address the three aforementioned challenges, with the goal of extending the applicability of characteristic mapping methods from idealized geometric flow problems to general physical settings involving arbitrary body forces and boundary conditions. Compared with the previous characteristic methods, our method differs in three aspects: 
%\bo{according to the three challenges I listed above, describe our three novelties as particle-based flow maps (also mentioning gradient-augmented P2G), path integral, and impulse form.}  
(1) Unlike existing methods that construct the flow map on Eulerian grids using high-order Hermite or spline interpolation (e.g., \cite{yin2021characteristic,yin2023characteristic,mercier2013characteristic,tessendorf2011characteristic}), we observe that particle trajectories themselves form an exact long-range flow map: each particle naturally carries its own mapping from the initial configuration to the current one, together with all transported physical quantities. This observation enables characteristic mapping to be coupled seamlessly with hybrid particle–grid solvers with gradient-augmented P2G (e.g., \cite{jiang2015affine,jiang2017angular}), allowing physical quantities and their gradients to be transported through the long-range flow map on particles and enabling our method to achieve higher accuracy at comparable algorithmic complexity.  (2) Most existing characteristic-mapping formulations rely on the vorticity form of the inviscid Euler equations. Although directly transporting vorticity helps preserve vortical structures, practical use requires nonlocal vorticity-to-velocity inversion (e.g., Biot-Savart), which is incompatible with standard velocity-based solvers (e.g. \cite{staniforth1991semi,stam2023stable,maccormack2002effect}) and makes it difficult for characteristic mapping to operate directly on the velocity field. Our key observation is that, while the velocity field itself does not satisfy a characteristic-mapping form due to incompressibility constraints, its gauge-transformed counterpart, the impulse \cite{cortez1995impulse,cortez1996impulse,ImpulseThesis}, does satisfy a clean characteristic evolution law. Moreover, impulse retains the same differential structure as velocity, allowing an impulse-based characteristic map to integrate naturally into conventional velocity-based solvers while avoiding the cost of vorticity inversion.  (3) Using the 1-form representation of the impulse equation, we rewrite the Navier–Stokes equations in a differential-form integral formulation that separates the evolution of impulse into (i) a component that is transported exactly through geometric mapping, and (ii) a component contributed by viscosity and body forces that can be evaluated through a path integral along particle trajectories. By leveraging the bidirectional flow map naturally provided by particles, our method computes and accumulates these path-integral contributions stably. This extension enables the impulse-based characteristic map, which was initially applicable only to inviscid Euler flows, to handle viscous and force-driven Navier–Stokes dynamics.  
%Due to the central role that particles play in our mapping scheme, we refer to our method as the \textbf{Particle Flow Map (PFM)} method.  
%\bo{Next, discuss these three aspects and their related literature in three paragraphs.}

%\bo{Talk about particle representation. I directly copied this paragraph from Zhiqi's draft. It is a good fit but requires some tailoring efforts.}
%The core idea of our approach is to construct the long-range flow map between frames on a set of moving particles while simultaneously transporting geometric quantities and integrating dissipative forces along these particles. 
Particle-based representations form the foundation of many Eulerian–Lagrangian hybrid solvers, where particles serve as carriers of advected quantities while Eulerian grids handle pressure and viscosity. This idea dates back to the Particle-in-Cell (PIC) method \cite{harlow1964particle}, which initiated the use of particles to reduce numerical diffusion in advection. Subsequent developments, including FLIP \cite{brackbill1986flip}, APIC \cite{jiang2017angular}, XPIC \cite{hammerquist2017new}, EPIC \cite{frey2022epic}, and the fifth-order HOPIC scheme \cite{edwards2012high}, systematically improved particle-to-grid transfer accuracy using increment schemes and higher-order moment transfer. More recent approaches, such as PolyPIC \cite{fu2017polynomial} and unified MLS-based formulations \cite{hu2018moving}, further enhance accuracy while reducing the reliance on complex interpolation kernels.  Despite these advances, existing particle–grid methods use particles only to perform short-step advection: particles move over a single time step and then immediately receive updated velocities from the grid. As a result, they do not exploit the fact that particle trajectories naturally encode long-range characteristic mapping information. The mapping they carry is repeatedly overwritten rather than accumulated, and repeated P2G/G2P transfers inevitably introduce dissipation. 

Recently, researchers have attempted to simulate vortical fluid flows by using particle flow maps to approximate long-range characteristic mappings of the impulse variable (e.g., \cite{zhou2024eulerian,li2024particle,chen2024solid} 
%\bo{more paper to cite Done}
). The core idea of the particle flow map method is to construct the long-range flow map between frames on a set of moving particles while simultaneously transporting geometric quantities and integrating dissipative forces along these particles. These particle–flow–map approaches have demonstrated strong capability in preserving the evolution of vortical structures over long time horizons and have produced visually compelling flow phenomena across various visual computing tasks (e.g., \cite{chen2025fluid, wang2025vpfm,wang2025cirrus,li2025clebsch,sun2025leapfrog,li2025adjoint} %\bo{cite more PFM works here Done}
). However, applying the particle flow map methodology to the full Navier–Stokes equations with rigorous treatment of all force terms and boundary conditions still remains unresolved. Moreover, its accuracy, convergence properties, and computational efficacy have not yet been validated through systematic benchmark tests on vortical flow problems for computational physics tasks. As a result, the current impact of particle flow maps is largely confined to visual simulation in computer graphics, while their potential for facilitating rigorous computational physics applications remains an open research question.

This paper adopts the perspective of the particle flow map method \cite{zhou2024eulerian}, which treats particles as exact carriers of the long-term flow map, and extends this paradigm from inviscid transport to the full Navier–Stokes equations with viscosity and external forcing. We briefly summarize the framework as follows: 
%\bo{Can we repurpose this paragraph more to this purpose(summarizing the framework)? Put more focus on treating viscosity and external forces. zhiqi Done before} 
By coupling this particle-based mapping with the impulse formulation, our method transports geometric quantities directly from the initial frame to the current frame without relying on repeated P2G/G2P transfers. This yields a characteristic-mapping framework with accuracy comparable to Hermite-interpolation–based characteristic mapping, but without requiring high-order gradient computations on the grid. In effect, our approach can be viewed as a long-range, non-dissipative generalization of classical particle-grid advection schemes. While particle trajectories provide the long-range geometric mapping required for characteristic transport, one must still identify a physical quantity whose evolution is compatible with such a mapping. For incompressible flows, the velocity field does not satisfy a characteristic form because the divergence-free constraint prevents direct mapping of velocity from the initial state. A natural solution comes from the classical impulse formulation, originally introduced by Buttke \cite{buttke1992lagrangian} and later developed by Cortez and collaborators \cite{cortez1995impulse,cortez1996impulse,cortez1998accuracy,cortez2000vortex} and others \cite{buttke1993velicity,buttke1993turbulence,summers2000representation}. In this formulation, the velocity is expressed via a gauge transformation  \( \mathbf{m} = \mathbf{u} + \nabla \phi \) where the impulse $\mathbf{m}$ can be mapped along characteristics without being constrained by incompressibility. Prior impulse-based methods typically used Lagrangian tracking to update $m$, but treated it only as a particle state rather than as a geometric quantity that can be mapped consistently through a long-range flow map.  In contrast, we treat impulse as the primary geometric variable carried by particles, allowing it to be mapped exactly by our PFM construction.  The velocity field is subsequently recovered through a standard incompressibility projection, thereby avoiding the difficulty of reconstructing the harmonic component of the velocity inherent in vorticity-based characteristic mapping approaches \cite{mizukami1983stream,tezduyar1988petrov,10.1145/3592402}.  
Crucially, extending inviscid Euler flow to general Navier-Stokes flow requires incorporating dissipative forces and body forces, which cannot be captured through pure geometric transport. To address this, we employ the 1-form representation of the impulse equation, which expresses the Navier–Stokes equations as a differential-form integral form. This formulation naturally separates the impulse update into two mapped components:(i) a geometric mapping component, which is mapped exactly along the particle-based long-range flow map; and (ii) a non-geometric component, including viscosity and body forces, represented by a path integral mapped along particle trajectories.  By leveraging the bidirectional particle flow map provided by PFM, these path-integral contributions can be computed and accumulated robustly over long time intervals. This formulation extends impulse-based characteristic mapping, which is previously restricted to inviscid Euler equations (e.g., \cite{yin2021characteristic,yin2023characteristic}), to general viscous and force-driven Navier–Stokes dynamics in a manner compatible with standard velocity-based solvers. Together, the particle flow map, path-integral formulation, and impulse-based gauge form collectively form the foundation of our characteristic-mapping framework.

%\bo{A paragraph to summarize the main contributions and introduce the structure of the rest of the paper.}
The main contributions of this work are summarized as follows:
\begin{enumerate}
    \item A new representation of characteristic mapping is introduced, in which long-range flow mappings are carried directly by particle trajectories rather than reconstructed on Eulerian grids. This particle-based formulation removes the need for high-order Hermite interpolation and supports gradient-augmented particle-to-grid transfers for improved geometric accuracy.
    
    \item The impulse field is identified as the primary variable admitting a long-range characteristic mapping while retaining structural compatibility with velocity-based incompressible solvers. This removes the need for nonlocal vorticity–velocity inversion inherent in vorticity-based characteristic mapping methods.
    
    \item Through the 1-form representation of the impulse equation, the Navier–Stokes equations are recast into an integration formula that naturally separates impulse evolution into a geometric mapping component and a non-geometric component. The latter, arising from viscosity and body forces, is evaluated through path integrals along particle trajectories and accumulated consistently over long time intervals.
    
    \item By combining particle flow maps, impulse mapping, and trajectory-based path integrals within a hybrid particle–grid computational structure, a complete solver called the Particle Flow Map (PFM) is constructed, extending characteristic mapping techniques beyond inviscid Euler flows to viscous, force-driven, and boundary-influenced Navier–Stokes problems.
\end{enumerate}

The remainder of this paper is organized as follows. Section~\ref{sec:physics_model} first presents the impulse formulation of the Navier–Stokes equations, and Section~\ref{sec:foundation_flow_map} reviews the fundamental concepts of characteristic mappings and flow maps. Building on these preliminaries, Section~\ref{sec:solution_inviscid_impulse} establishes the characteristic-mapping property of the impulse variable and derives its analytical solution for inviscid Euler flows. Section~\ref{sec:particle_flow_map} then introduces the particle-based representation of flow maps from which our method is derived, and Section~\ref{sec:pfm_based_calculation} uses this representation to compute impulse transport through particle carried flow maps. Section~\ref{sec:solution_impulse} subsequently extends this formulation to Navier–Stokes flows involving viscosity, body forces, and solid boundaries via a trajectory-based path-integral approach, and Section~\ref{sec:path_integration_buffer} introduces the flow map formed calculation of these path integrals on particles, which together with impulse mapping forms the complete solver of our method. Section~\ref{sec:validation} and \ref{sec:result} finally evaluate the method through a series of numerical experiments. %\bo{Avoid using the term PFM as much as we can. Done}

\section{Physics Model}\label{sec:physics_model}
We start with the incompressible Navier–Stokes equations in $d$-dimensional space with uniformly constant density as:
\begin{equation}\label{eq:NS_equation}
    \left\{
    \begin{aligned}
        &(\partial_t+\mathbf{u}\cdot\nabla)\mathbf{u}  = -\nabla p +\nu\Delta \mathbf{u}+ \mathbf{f},\\
        &\nabla\cdot \mathbf{u} =0,
    \end{aligned}
    \right.
\end{equation}
where $\nu$ is the kinematic viscosity and $\mathbf{f}$ represents external forces such as gravity or coupling forces. Following the impulse-form Navier-Stokes equations \cite{ImpulseThesis,cortez1995impulse,cortez1996impulse,cortez1998accuracy,cortez2000vortex}, we define a gauge transformation of $\mathbf{u}$ as $\mathbf{m} = \mathbf{u} + \nabla \phi$, where $\phi$ is a scalar field, and $\mathbf{m}$ is referred to as the impulse variable. The Navier–Stokes equations \eqref{eq:NS_equation} can then be expressed in terms of $\mathbf{m}$ as:
\begin{equation}\label{eq:impulse_equation}
\left\{
    \begin{aligned}
        &(\partial_t + \mathbf{u} \cdot \nabla) \mathbf{m} = -(\nabla \mathbf{u})^\mathrm{T} \mathbf{m} + \nu \Delta \mathbf{m} + \mathbf{f},\\
        &\Delta \phi = \nabla \cdot \mathbf{m},\\
        &\mathbf{u} = \mathbf{m} - \nabla \phi,
    \end{aligned}
    \right.
\end{equation}
where $\phi$ also satisfies the advection equation 
\[
\frac{D\phi}{Dt} + \frac{1}{2} \|\mathbf{u}\|^2 - p - \nu \Delta \phi = 0.
\]
In practice, $\phi$ can be calculated directly by solving the Poisson equation $\Delta \phi = \nabla \cdot \mathbf{m}$, as $\mathbf{m} = \mathbf{u} + \nabla \phi$ represents the Helmholtz decomposition of $\mathbf{m}$ with respect to the divergence-free velocity field $\mathbf{u}$.

%\bo{I am not sure if we should put these paragraphs here because they are related to numerical method rather than physical model. Move them to somewhere else? Not sure where to put since this has to be before Sec.3 since it envolves why we have 3.2 and after that having 4.1. It also has to be after Sec 2 since that's the first place where we have advection equation for impulse.}
In this paper, we develop a time-split framework to solve the Navier-Stokes equation in terms of impulse (\ref{eq:impulse_equation}), thereby solving the Navier-Stokes equation (\ref{eq:NS_equation}). In our time-split framework, equation (\ref{eq:impulse_equation}) is divided into three steps. First, by employing the particle flow map method introduced in section \ref{sec:particle_flow_map}, we accurately solve the advection equation of impulse without the viscous term and external force term in section \ref{sec:pfm_based_calculation}:
\begin{equation}\label{eq:invisid_impulse_advection_equation}
    (\partial_t + \mathbf{u} \cdot \nabla) \mathbf{m} = -(\nabla \mathbf{u})^\mathrm{T} \mathbf{m}.
\end{equation}
In the second step, the viscous forces $\nu \Delta \mathbf{m}$ and external forces $\mathbf{f}$ are solved with a path integration buffer compatible with the flow map approach used in the first step, which is introduced in section \ref{sec:path_integration_buffer}. Finally, we obtain the velocity \( \mathbf{u} \) from $\mathbf{m}$ through a divergence-free projection \( \mathbf{u} = \mathbb{P}\mathbf{m} \) in section \ref{sec:pfm_based_calculation}, where $\mathbb{P}$ is the projection operator. It is worth noting that, our method can also be applied to solve the inviscid Euler equation with only the first and third steps.

\section{Flow Map-Based Computation of Inviscid Impulse}
\begin{table}
    \centering
    \scriptsize
    \caption{\textbf{Notation Table}}
    \begin{tabular}{|c|c|c||c|c|c|}
        \hline
        Notation & Type & Definition & Notation & Type & Definition \\
        \hline
        $*_t$ & - & Quantity defined at time $t$ & $\mathbf{x}$ & vector & Position \\
        \hline
        $\mathbf{X}$ & Vector & Backward flow map & $\mathbf{Y}$ & Vector & Forward flow map  \\
        \hline
        $\mathbf{T}$ & Matrix & Backward flow map Jacobian & $\mathbf{F}$ & Matrix & Forward flow map Jacobian  \\
        \hline
        $\mathbf{u}$ & Vector & Velocity & 
        $\mathbf{m}$ & Vector & Impulse Variable\\
        \hline
        $*_\alpha$ & - & Quantity stored on grids $t$ & \hspace{12pt}$*_p$ & - & Quantity stored on particles  \\
        \hline
        $\mathbf{\Gamma}$ & Vector & Integration Buffer &
        $\mathbf{A}$ & Matrix & Affine matrix \\
        \hline
    \end{tabular}
    \label{tab:notation}
\end{table}
\subsection{Foundation of Flow Map}\label{sec:foundation_flow_map}
% \bo{Need a figure}
%\bo{A comment on the notation. See my notes. Done}
From initial time $s$, consider the fluid domain at arbitrary time $t$ as $U_t, t\ge s$.  For arbitrary divergence-free velocity field $\mathbf{u}: \mathbb{R}^d\times \mathbb{R}_+ \to \mathbb{R}^d$,  the backward flow map is defined as a one-parameter family of maps $\mathbf{X}_t: U_t\to U_s$, satisfying
\begin{equation}\label{eq:flow_map_advection_equation}
\left\{
\begin{aligned}
        &(\partial_t  + \mathbf{u}\cdot \nabla )\mathbf{X}_t = 0,\\
        &\mathbf{X}_s(\mathbf{x}) = \mathbf{x},
\end{aligned}
\right.
\end{equation}
and the forward flow map $\mathbf{Y}_t: U_s\to U_t$ is defined as the inverse map of the backward flow map $\mathbf{Y}_t = \mathbf{X}_t^{-1}$.  Forward flow maps $\mathbf{Y}_t$ have $\mathbf{u}_t$ as the velocity of their flow
\begin{equation}\label{eq:forward_flow_map_advection_equation}
    \partial_t \mathbf{Y}_t(\mathbf{x}) = \mathbf{u}_t(\mathbf{Y}_t(\mathbf{x})).
\end{equation}
\begin{figure}
    \centering
    \includegraphics[width=0.8\linewidth]{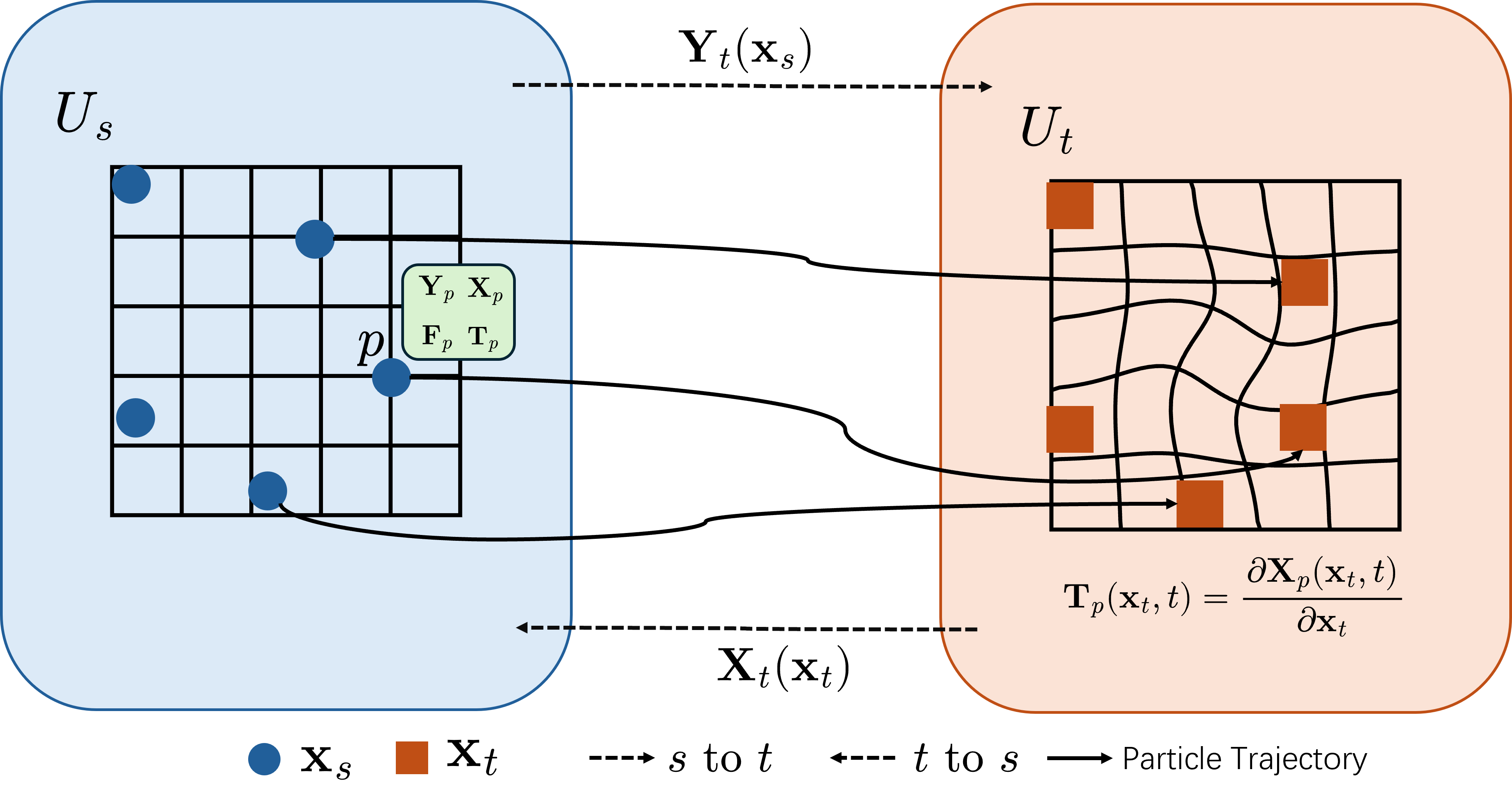}
    \caption{Illustration for flow map and flow map Jacobians}
    \label{fig:flowmap_illu}
\end{figure}
In previous literature \cite{nave2010gradient,mercier2013characteristic,yin2021characteristic,yin2023characteristic}, the backward flow map $\mathbf{X}_t$ is also called the characteristic map, since for any \( \mathbf{\gamma}_0 \in U_s \), \( \mathbf{\gamma}(t) = \mathbf{X}^{-1}_t(\mathbf{\gamma}_0) \) represents a characteristic curve associated with the velocity \( \mathbf{u}\).  

The Jacobians of the backward and forward flow maps are defined as $\mathbf{T}(\mathbf{x},t) = \frac{\partial \mathbf{X}(\mathbf{x},t)}{\partial \mathbf{x}}, \mathbf{x}\in U_t$ and $\mathbf{F}(\mathbf{x},t) = \frac{\partial \mathbf{Y}(\mathbf{x},t)}{\partial \mathbf{x}}, \mathbf{x}\in U_s$ respectively, with following relationships hold
\begin{equation}
\left\{
 \begin{aligned}
 \mathbf{F}(\mathbf{X}_t(\mathbf{x}),t)\mathbf{T}_t(\mathbf{x},t) &= \mathbf{I}(\mathbf{x}), \mathbf{x}\in U_t,   \\
 \mathbf{F}(\mathbf{x},t)\mathbf{T}(\mathbf{Y}_t(\mathbf{x}),t) &= \mathbf{I}(\mathbf{x}), \mathbf{x}\in U_s,     
 \end{aligned}
 \right.
\end{equation}
where \(\mathbf{I}\) represents the identity matrix field on \(U_s\) or \(U_t\).  By differentiating both sides of the equation (\ref{eq:flow_map_advection_equation}), the 
evolution equation for \(\mathbf{T}(\mathbf{x},t)\) can be directly obtained
\begin{equation}
\left\{
\begin{aligned}
        &(\partial_t  + \mathbf{u}\cdot \nabla )\mathbf{T}_t = -\mathbf{T}_t \nabla \mathbf{u},\\
        &\mathbf{T}_s(\mathbf{x}) = \mathbf{I}(\mathbf{x}),
\end{aligned}
\right.
\end{equation}

Backward flow maps or characteristic maps have often been used in previous work \cite{yin2021characteristic,mercier2013characteristic,nave2010gradient,kolomenskiy2016adaptive} to compute scalar fields $\phi(\mathbf{x},t)$ at arbitrary time $t$ from the initial condition directly and accurately, using traced backward flow maps $\mathbf{X}_t$, based on the analytic solution $\phi(\mathbf{x},t)=\phi(\mathbf{X}_t(\mathbf{x}),0)$ to the advection equation 
\begin{equation}
\left\{
    \begin{aligned}
        &(\partial_t+\mathbf{u}\cdot \nabla)\phi =0,\\
        &\phi(\mathbf{x},0) = \phi_0(\mathbf{x}),     
    \end{aligned}
    \right.
\end{equation}
where \(\phi\) can be a 2D vorticity field \cite{yin2021characteristic} or any level set field \cite{mercier2013characteristic,nave2010gradient,kolomenskiy2016adaptive}. In this paper, we will use the analytic solution of inviscid impulse advection represented by the flow map to accurately compute the impulse $\mathbf{m}$. Additionally, we provide a path integrator buffer based on the flow map to handle the viscous and external force terms in the impulse advection, ultimately solving the Navier-Stokes equation through the flow map.

\subsection{Solution to Inviscid Impulse Advection}\label{sec:solution_inviscid_impulse}
By deriving the analytical solution of the inviscid impulse advection equation (\ref{eq:invisid_impulse_advection_equation}), \cite{buttke1992lagrangian, ImpulseThesis} proposes the following mapping formula for the impulse $\mathbf{m}$:
\begin{equation}\label{eq:solution_invisid_impulse}
\left\{
\begin{aligned}
    \mathbf{m}(\mathbf{x},t) &= \mathbf{T}^\top(\mathbf{x},t) \mathbf{m}(\mathbf{X}_t(\mathbf{x}),0), \\
    \mathbf{m}(\mathbf{x},0) &= \mathbf{u}_0(\mathbf{x}),
\end{aligned}
\right.
\end{equation}  
where \(\mathbf{u}_0\) denotes the initial velocity field. In our method, equation (\ref{eq:solution_invisid_impulse}) is used to accurately and directly compute the advection of \(\mathbf{m}\) from the initial velocity field after the flow map \(\mathbf{X}_t\) and its Jacobians \(\mathbf{T}_t\) are calculated.

In \cite{ImpulseThesis}, equation (\ref{eq:solution_invisid_impulse}) is presented based on the concept of "surface elements", and the equivalence between equation (\ref{eq:solution_invisid_impulse}) and equation (\ref{eq:invisid_impulse_advection_equation}) is verified by taking the partial derivative of equation (\ref{eq:solution_invisid_impulse}) to time. This method of formulation cannot be naturally extended to the viscous case, and for convenience in deriving the path integrator buffer for viscous and external force terms in Section \ref{sec:viscous_case}, we provide an alternative derivation of Equation (\ref{eq:solution_invisid_impulse}) using the language of differential forms, which can be easily extended to the viscous formulation. 
 We begin by rewriting equations (\ref{eq:solution_invisid_impulse}) and (\ref{eq:invisid_impulse_advection_equation}) in terms of the language of differential forms and show that equation (\ref{eq:solution_invisid_impulse}) is a natural result of the property of Lie derivative
\begin{equation}\label{eq:Lie_property}
\frac{\partial}{\partial t} \mathbf{\Phi}_t^* \mathsf{T}_t = \mathbf{\Phi}_t^* \left( \frac{\partial}{\partial t} + \mathcal{L}_{\mathbf{u}} \right) \mathsf{T}_t,    
\end{equation}
where $\mathbf{\Phi}_t$ is an arbitrary one-parameter family of maps that $\mathbf{\Phi}_t: U_s\to U_t$ with $\mathbf{u}$ as their velocity of the flow, the superscript asterisk $^*$ denotes the pullback of the map $\mathbf{\Phi}_t$, and $\mathsf{T}_t$ is a time-dependent $k$-form in $U_t$ for any integer $k\ge0$

Define the impulse 1-form $\mathsf{m}_t = \mathbf{m}_t^\flat$ for any $t \ge s$, where $\flat$ is the musical isomorphism operator that turns a vector field into a $1$-form.  The inviscid impulse advection equation (\ref{eq:invisid_impulse_advection_equation}) is equivalent to the Lie-transport equation of the impulse 1-form $\mathsf{m}$:
\begin{equation}\label{eq:invisid_impulse_Lie_advection_equation}
    (\partial_t+\mathcal{L}_{\mathbf{u}})\mathsf{m}_t = 0,
\end{equation}
In $\mathbb{R}^d$, the Lie derivative of the impulse 1-form \( \mathsf{m} \) can be computed as \( \mathcal{L}_{\mathbf{u}} \mathsf{m} = i_\mathbf{u} d\mathsf{m} + d(i_\mathbf{u} \mathsf{m}) \) by Cartan's formula, and finally expressed in vector form in $\mathbb{R}^d$ as \((\mathbf{u} \cdot \nabla)\mathbf{m} + (\nabla \mathbf{u})^\top \mathbf{m} \),  which proves the equivalence between equation (\ref{eq:invisid_impulse_advection_equation}) and (\ref{eq:invisid_impulse_Lie_advection_equation}). 

Equation (\ref{eq:invisid_impulse_Lie_advection_equation}) essentially states that the impulse 1-form \( \mathsf{m} \) is purely Lie-advected by the velocity field \( \mathbf{u} \).  After applying $\mathbf{Y}_t^*$ to both sides of equation \ref{eq:invisid_impulse_Lie_advection_equation}, by the property (\ref{eq:Lie_property}), we have $\partial_t(\mathbf{Y}_t^* \mathsf{m}) = 0$.  This provides an expression for the impulse 1-form as the pullback of the initial condition by the backward flow map \(\mathbf{X}_t \):
\begin{equation}\label{eq:solution_lie_invisid_impulse}
    \mathsf{m}_t = \mathbf{X}_t^* \mathsf{m}_0.
\end{equation}
For the impulse 1-form \( \mathsf{m} \) defined in \( \mathbb{R}^d \), the pullback in Equation (\ref{eq:solution_lie_invisid_impulse}) can be calculated as \( \mathbf{m}_t(x) = \nabla \mathbf{X}_t^\top \mathbf{m}_0(\mathbf{X}_t(\mathbf{x})) \) in corresponding vector form, which essentially provides Equation (\ref{eq:solution_invisid_impulse}). 

\section{Numerical Calculation}
\subsection{Particle Flow Map Method} \label{sec:particle_flow_map}
As mentioned earlier, we aim to develop a hybrid grid-particle method algorithm, based on the idea that particles themselves represent the flow map, performing computations using a set of particles and a Cartesian grid. The subscript $_p$ is used to denote any quantity carried by particle $p \in \mathbb{P} $, where \( \mathbb{P} \) is the set of all participating particles, and the subscript $_\alpha$ denotes any quantity on the grid node $\alpha \in \mathbb{A}$, where \( \mathbb{A} \) is the set of all grid nodes.  We use \( \mathbf{x}_p(t) \) to represent the position of particle $p$ at time $t$, \( \mathbf{x}_\alpha \) to represent the position of grid node $\alpha$, and \( \mathbf{u}_\alpha(t) \) to represent the discretized fluid velocity field on the grid.  Like other hybrid methods, such as the PIC (Particle-in-Cell Method) or FLIP (Fluid Implicit Particle Method), the position \( \mathbf{x}_p \) of particle $p$ is updated as
\begin{equation}\label{eq:advection_position_particles}
    \frac{d}{dt}\mathbf{x}_p(t) = \mathbf{u}_t(\mathbf{x}_p(t)),
\end{equation}
The value of the velocity field at the particle position, \( \mathbf{u}_t(\mathbf{x}_p(t)) \), and its gradient matrix, \( \nabla\mathbf{u}_t(\mathbf{x}_p(t)) \), can be calculated via grid-to-particle interpolation using a quadratic kernel function $\omega(\cdot,\cdot)$:
\begin{equation}\label{eq:g2p_u}
\left\{
\begin{aligned}
     &\mathbf{u}_t(\mathbf{x}_p(t)) = \sum_{\alpha \in \mathbb{N}_p} \mathbf{u}_\alpha(t)\omega(\mathbf{x}_\alpha, \mathbf{x}_p), \\
    &\nabla \mathbf{u}_t(\mathbf{x}_p(t)) = \sum_{\alpha \in \mathbb{N}_p} \mathbf{u}_\alpha(t)\nabla_{\mathbf{x}_p}\omega(\mathbf{x}_\alpha, \mathbf{x}_p) ,
\end{aligned}
\right.
\end{equation}
where \( \mathbb{N}_p = \{ \alpha \in \mathbb{A} \mid \omega(\mathbf{x}_\alpha, \mathbf{x}_p) > 0 \} \) represents the set of neighboring grid nodes for particle $p$.

In previous characteristic mapping literature \cite{yin2021characteristic,mercier2013characteristic,nave2010gradient,kolomenskiy2016adaptive}, the backward flow map $\mathbf{X}_t$ was tracked on grids through the GALS (Gradient Augmented Level Set) framework \cite{nave2010gradient} and developed into a corresponding grid-based method. However, we observe that in hybrid methods, the particles that move with the fluid inherently constitute the forward and backward flow map themselves
\begin{equation}\label{eq:particles_are_flow_maps}
\left\{
    \begin{aligned}
        \mathbf{x}_p(t) &= \mathbf{X}_t(\mathbf{x}_p(0)), \\
        \mathbf{x}_p(0) &= \mathbf{Y}_t(\mathbf{x}_p(t)),    
    \end{aligned}    
    \right.
\end{equation}
Namely, \( \mathbf{x}_p(t) \) is the forward flow map of \( \mathbf{x}_p(0) \), while \( \mathbf{x}_p(0) \) is the backward flow map of \( \mathbf{x}_p(t) \).  Equation (\ref{eq:particles_are_flow_maps}) holds because the particle motion equation (\ref{eq:advection_position_particles}) is equivalent to the evolution equation (\ref{eq:forward_flow_map_advection_equation}) of the forward flow map. In addition to the flow map itself, the Jacobians of the backward and forward flow maps can also be carried and evolved by the motion of particles with the velocity field \(\mathbf{u}\) on the background grid. Denoting the Jacobians of the backward and forward flow maps carried by the particles as \( \mathbf{T}_p(t) = \mathbf{T}(\mathbf{x}_p(t), t) \) and \( \mathbf{F}_p(t) = \mathbf{F}(\mathbf{x}_p(0), t) \), respectively, it can be proved that \( \mathbf{T}_p \) and \( \mathbf{F}_p \) follow the advection equation as:
\begin{equation} \label{eq:advection_FT_on_particles}
    \left\{
    \begin{aligned}
        \frac{d}{dt}\mathbf{F}_p(t) &= \nabla \mathbf{u}(\mathbf{x}_p(t),t)  \mathbf{F}_p(t), \\
        \frac{d}{dt} \mathbf{T}_p(t) &= -\mathbf{T}_p(t) \nabla \mathbf{u}(\mathbf{x}_p(t),t),
    \end{aligned}
    \right.
\end{equation}
which can be directly derived by taking the time derivative of the definitions of \( \mathbf{T}_p \) and \( \mathbf{F}_p \).

In each substep, we compute equation (\ref{eq:advection_position_particles}) to move the particles, while simultaneously computing equation (\ref{eq:advection_FT_on_particles}) to evolve \( \mathbf{F}_p \) and \( \mathbf{T}_p \) to the next time step, using the 4th-order Runge-Kutta method and equation (\ref{eq:g2p_u}) to calculate $\mathbf{u}(\mathbf{x}_p(t),t)$ and $\nabla \mathbf{m}(\mathbf{x}_p(t),t)$. This approach allows us to efficiently and accurately track the backward and forward flow maps and their Jacobians without relying on the complex Hermite interpolation and time integration schemes used in the grid-based GALS framework\cite{nave2010gradient}.  Since we use particles to represent and track flow maps and their Jacobians, we call this hybrid framework of calculating flow maps as the particle flow map method, abbreviated as PFM.

\subsection{Numerical Calculation based on Particle Flow Map}\label{sec:pfm_based_calculation}
In this section, we introduce how to perform numerical computations based on the aforementioned framework to complete step 1 and step 3 in section \ref{sec:physics_model}, forming a numerical algorithm for solving the inviscid Euler equation.

At time $t$, after the backward and forward flow maps and their Jacobians are obtained by the method in section \ref{sec:particle_flow_map},  with the initial velocity \( \mathbf{u}_{0,p} = \mathbf{u}_0(\mathbf{x}_p(0)) = \mathbf{m}_0(\mathbf{x}_p(0)) \) carried by particles, equation (\ref{eq:solution_invisid_impulse}) can be solved to obtain the value of the impulse field \( \mathbf{m}_p(t) = \mathbf{m}_t(\mathbf{x}_p(t), t) \) at the position $\mathbf{x}_p(t)$ of particle \( p \) 
\begin{equation}\label{eq:m_calculation}
    \mathbf{m}_p(t) = \mathbf{T}_p(t)^\top \mathbf{m}_{0,p}.
\end{equation}
The values \( \mathbf{m}_p(t) \) computed on the particles constitute a particle-based discretization of the field \( \mathbf{m}_t \), which completes the first step, solving inviscid impulse advection, mentioned in Section \ref{sec:physics_model}.

For inviscid fluids, only the third step is required afterward, where the impulse \( \mathbf{m}_t \) is projected into the velocity field. The projection of the impulse \( \mathbf{m}_t \) into the velocity field is performed on the grid to obtain the background grid velocity, which is used to evolve the particle positions and their flow map information. Before projection, \( \mathbf{m}_p(t) \) on the particles is interpolated to the grid through an APIC-style particle-to-grid (P2G) transfer process
\begin{equation}\label{eq:m_p2g}
    \mathbf{m}_\alpha(t) = \sum_{p \in N_\alpha} (\mathbf{m}_p(t)+\mathbf{A}_p(\mathbf{x}_\alpha - \mathbf{x}_p))\omega(\mathbf{x}_p,\mathbf{x}_\alpha),
\end{equation}
where \( \mathbb{N}_\alpha = \{ p \in \mathbb{P} \mid \omega(\mathbf{x}_p, \mathbf{x}_\alpha) > 0 \} \) represents the set of neighboring grid nodes for particle $p$ and the affine matrix \( \mathbf{A}_p \) is the gradient \( \nabla \mathbf{m}_{t-\Delta t}(\mathbf{x}_p(t)) \) at position $\mathbf{x}_p(t)$ of the impulse \( \mathbf{m} \) on the grid last time step, calculated using the same method as computing \( \nabla \mathbf{u}_t(\mathbf{x}_p(t)) \) in equation (\ref{eq:g2p_u}).  For projection, the Poisson equation \(\Delta \phi = \nabla \cdot \mathbf{m}\) is discretized into a linear equation on the grid. This equation is then effectively solved using the Multigrid Preconditioned Conjugate Gradient (MGPCG) \cite{mcadams2010parallel} method to obtain \(\phi_\alpha\). Finally, the fluid velocity at time \(t\) is obtained as \(\mathbf{u}_\alpha = \mathbf{m}_\alpha - \nabla \phi_\alpha\). This completes step 3 in Section \ref{sec:physics_model}.

\subsection{Reinitialization of Flow Map}
Due to numerical inaccuracies and limited representation resolution, flow map methods (as well as characteristic map methods \cite{yin2021characteristic, yin2023characteristic}) require reinitialization of the flow map and its Jacobian to prevent computational instability. In our method, reinitialization is performed by uniformly redistributing the particles, resetting the flow map Jacobians \( \mathbf{T}_p \) and \( \mathbf{F}_p \) to the identity matrix, and updating the initial time \( s \) in Equation (\ref{eq:solution_invisid_impulse}) to the current time.

Combining the above processes, we have developed a particle flow map–based algorithm for solving the inviscid Euler equations, which is summarized in Algorithm \ref{alg:pfm_euler}.

\section{Inclusion of Viscous and External Force}\label{sec:viscous_case}

In this section, we focus on step 2 of our time-split framework mentioned in section \ref{sec:physics_model}, which addresses the viscous term \(\nu \Delta \mathbf{m}\) and external force term \(\mathbf{f}\) in the Navier-Stokes equation (\ref{eq:impulse_equation}) in terms of impulse. First, in section \ref{sec:solution_impulse}, using the language of differential forms introduced in section \ref{sec:solution_inviscid_impulse}, We will derive the integral formulation corresponding to equation (\ref{eq:impulse_equation}), which is an extension of the inviscid analytic solution (\ref{eq:solution_lie_invisid_impulse}) and (\ref{eq:solution_invisid_impulse}). Then, in section \ref{sec:path_integration_buffer}, we will introduce how to compute the additional integral terms arising from viscous forces and external forces in the integral formulation using a path integration buffer compatible with our framework.

\subsection{Solution to Impulse Equation with Viscous and External Force}\label{sec:solution_impulse}

Similar to the inviscid counterpart, equation (\ref{eq:impulse_equation}) can be expressed as a differential equation for the impulse 1-form \(\mathsf{m}_t \) with the Lie derivative
\begin{equation}\label{eq:lie_impulse_equation}
    (\partial_t+\mathcal{L}_\mathbf{u}) \mathsf{m}_t = \nu (\Delta \mathbf{m})^\flat+ \mathbf{f}^\flat .
\end{equation}

Applying pullback operator $\mathbf{Y}_t^*$ t to both sides of equation (\ref{eq:lie_impulse_equation}), with property (\ref{eq:Lie_property}), we have:
\begin{equation}
    \partial_t (\mathbf{Y}_t^* \mathsf{m}_t) = \mathbf{Y}_t^* (\nu (\Delta \mathbf{m})^\flat+ \mathbf{f}^\flat).
\end{equation}
 Integrating the above equation over the time interval \( s \leq \tau \leq t \), we have an $1$-form-expressed integral formulation, which is an extension of the inviscid analytic solution (\ref{eq:solution_lie_invisid_impulse}):
\begin{equation}\label{eq:solution_lie_impulse}
    \mathsf{m}_t =\mathbf{X}_t^* \mathsf{m}_0 + \mathbf{X}_t^* \int_{s}^t\mathbf{Y}_\tau^* (\nu (\Delta \mathbf{m})^\flat+ \mathbf{f}^\flat) d\tau.
\end{equation}
After converting equation (\ref{eq:solution_lie_impulse}) into vector form, we have:
\begin{equation}\label{eq:solution_impulse}
\begin{aligned}
    \mathbf{m}(\mathbf{x},t) =&\mathbf{T}^\top(\mathbf{x},t) \mathbf{m}(\mathbf{X}_t(\mathbf{x}),0) \\
    &+ \mathbf{T}^\top(\mathbf{x},t) \int_{s}^t\mathbf{F}^\top(\mathbf{X}_t(\mathbf{x}),\tau) (\nu \Delta \mathbf{m}+ \mathbf{f})(\mathbf{Y}_\tau\circ\mathbf{X}_t(\mathbf{x}),\tau) d\tau .
\end{aligned}
\end{equation}
The formula above can be regarded as an extension of equation (\ref{eq:solution_invisid_impulse}), with an additional integral term caused by viscous forces and external forces. This formula serves as the basis for computing the impulse equation with viscous and external forces using the flow map. In the next section, we will introduce how to compute the integral terms in equation (\ref{eq:solution_impulse}) through the flow map.

\subsection{Path Integration Buffer Compatible with Particle Flow Map}\label{sec:path_integration_buffer}

Let $\mathbf{\Gamma}(\mathbf{x}, t)$ denote the additional integral term of viscous and external forces as $\mathbf{\Gamma}(\mathbf{x}, t) = \int_{s}^t \mathbf{F}^\top(\mathbf{X}_t(\mathbf{x}), \tau) (\nu \Delta \mathbf{m} + \mathbf{f})(\mathbf{Y}_\tau \circ \mathbf{X}_t(\mathbf{x}), \tau) \, d\tau$.  To compute and track \(\mathbf{\Gamma}(\mathbf{x}, t)\), we define the path integration buffer \(\mathbf{\Gamma}_p(t) = \mathbf{\Gamma}(\mathbf{x}_p(t), t)\) for each particle \(p\), which store and update the value of $\mathbf{\Gamma}$. Using the flow map information \(\mathbf{x}_p(t)\), \(\mathbf{T}_p(t)\), and \(\mathbf{F}_p(t)\) tracked in section \ref{sec:particle_flow_map}, \(\mathbf{\Gamma}_p\) can be expressed as  
\begin{equation}\label{eq:path_integrator_buffer}
        \mathbf{\Gamma}_p(t) = \int_s^t \mathbf{F}^\top_p(t)(\nu\Delta \mathbf{m} + \mathbf{f})(\mathbf{x}_p(\tau),\tau)d\tau, 
\end{equation}
which essentially integrates \(\mathbf{F}_p^\top(\nu \Delta \mathbf{m} + \mathbf{f})\) along the trajectory $\mathbf{\gamma}_p$, where $\mathbf{\gamma}_p$ denotes the trajectory of particle \(p\), parameterized by time \(t\), as \(\mathbf{\gamma}_p(t) = \mathbf{x}_p(t)\). Therefore, we refer to \(\mathbf{\Gamma}_p\) as the path integration buffer.  On particles, the complete expression for impulse is:  
\begin{equation}\label{eq:impulse_with_path_integrator_update}
        \mathbf{m}_p(t) = \mathbf{T}_p(t)^\top \mathbf{m}_{0,p} + \mathbf{T}_p^\top(t)\mathbf{\Gamma_p(t)}.
\end{equation}

To calculate \(\mathbf{\Gamma}_p(t)\), it only needs to be updated at each time step by accumulating the viscous forces and external forces computed at current time step:
\begin{equation}\label{eq:path_integrator_update}
\mathbf{\Gamma}_p(t) = \mathbf{\Gamma}_p(t') + \int_{t'}^t \mathbf{F}^\top_p(t)(\nu\Delta \mathbf{m} + \mathbf{f})(\mathbf{x}_p(\tau),\tau)d\tau,
\end{equation}
where \(t' = t - \Delta t\) represents the time of the previous time step.  The term \((\nu\Delta \mathbf{m} + \mathbf{f})(\mathbf{x}_p(\tau),\tau)\) is calculated in a manner same as the interpolation process of \(\mathbf{u}_t(\mathbf{x}_p(t))\) described in equation (\ref{eq:g2p_u}). The integration in equation (\ref{eq:path_integrator_update}) is computed using a simple forward Euler method in our implementation.  After updating \(\mathbf{\Gamma}_p(t)\), according to equation (\ref{eq:impulse_with_path_integrator_update}), we combine \(\mathbf{\Gamma}_p(t)\) with the inviscid advected impulse \(\mathbf{m}^{\text{inviscid}}_{p}(t) = \mathbf{T}_p(t)^\top \mathbf{m}_{0,p}\), computed in section \ref{sec:pfm_based_calculation}, to obtain the final impulse \(\mathbf{m}_p(t)\) at time \(t\). This completes step 2 outlined in section \ref{sec:physics_model}.

By combining the computation of step 2 discussed in this section with the step 1 and step 3 computation described in section \ref{sec:pfm_based_calculation}, we have developed a Particle Flow Map–based algorithm for solving the Navier-Stokes equation (\ref{eq:impulse_equation}) with viscous and external force terms. The algorithm is summarized in Algorithm \ref{alg:pfm_ns}.

\begin{algorithm}[t]
\caption{Particle Flow Map Method for Euler Equations}
\label{alg:pfm_euler}
\textbf{Initialize:} $\mathbf{u}$ to initial velocity; $\mathbf{T}_p(\mathbf{x})$ to $\mathbf{I}$
\begin{algorithmic}[1]
\For{$k$ in total steps}
\State $i \gets k \mod {n}$;
\If{$i$ = 0}
\State Uniformly distribute particles;

\State Reinitialize $\mathbf{m_0}$ ; \hfill $\triangleright$ Eq.~(\ref{eq:solution_invisid_impulse})
\State Reinitialize $\mathbf{T}_p(\mathbf{x})$ to $\mathbf{I}$.
\EndIf

\State Compute $\Delta t$ with $\mathbf{u}_{\alpha}(t)$ and the CFL number;
\State Estimate grid midpoint velocity $\mathbf{u}_{\alpha}(t + \frac{1}{2}\Delta t)$ using RK4 integration of $\mathbf{u}(t)$.
\State March $\mathbf{x}_p(t)$, $\mathbf{T}_p(\mathbf{x})$ with $\mathbf{u}_{\alpha}(t + \frac{1}{2}\Delta t)$ and $\Delta t$ using RK4; \hfill $\triangleright$ Eq.~(\ref{eq:advection_FT_on_particles})

\State Compute $\mathbf{m}_p(t)$ with $\mathbf{m}_{0,p}$ and $\mathbf{T}_p$; \hfill $\triangleright$ Eq.~(\ref{eq:m_calculation})

\State Compute $\mathbf m_{\alpha}$ by a APIC style P2G using $\mathbf m_{p}$; \hfill $\triangleright$ Eq.~(\ref{eq:m_p2g})
\State Compute $\mathbf{u}_{\alpha}$ by solving $\textbf{Poisson}(\mathbf{m}_{\alpha})$;
\EndFor{}
\end{algorithmic}
\end{algorithm}

\begin{algorithm}[t]
\caption{Particle Flow Map Method for NS Equations}
\label{alg:pfm_ns}
\textbf{Initialize:} $\mathbf{u}$ to initial velocity; $\mathbf{T}_p(\mathbf{x})$, $\mathbf{F}_p(\mathbf{x})$ to $\mathbf{I}$
\begin{algorithmic}[1]
\For{$k$ in total steps}
\State $i \gets k \mod {n}$;
\If{$i$ = 0}
\State Uniformly distribute particles;
\State Reinitialize $\mathbf{m_0}$ ; \hfill $\triangleright$ Eq.~(\ref{eq:solution_invisid_impulse})
\State Reinitialize $\mathbf{T}_p(\mathbf{x})$, $\mathbf{F}_p(\mathbf{x})$ to $\mathbf{I}$;
\State Set $\mathbf{\Gamma}_p(t)$ to 0.
\EndIf

\State Compute $\Delta t$ with $\mathbf{u}_{\alpha}(t)$ and the CFL number;
\State Estimate grid midpoint velocity $\mathbf{u}_{\alpha}(t + \frac{1}{2}\Delta t)$ using RK4 integration of $\mathbf{u}(t)$.
\State March $\mathbf{x}_p(t)$, $\mathbf{T}_p(\mathbf{x})$, $\mathbf{F}_p(\mathbf{x})$ with $\mathbf{u}_{\alpha}(t + \frac{1}{2}\Delta t)$ and $\Delta t$ using RK4; \hfill $\triangleright$ Eq.~(\ref{eq:advection_FT_on_particles})

\State Compute $\mathbf{m}_p(t)$ with $\mathbf{m}_{0,p}$, $\mathbf{\Gamma}_p(t)$ and $\mathbf{T}_p$; \hfill $\triangleright$ Eq.~(\ref{eq:impulse_with_path_integrator_update})

\State Compute $\mathbf m_{\alpha}$ by a APIC style P2G using $\mathbf m_{p}$; \hfill $\triangleright$ Eq.~(\ref{eq:m_p2g})

\State Accumulate external force and viscosity force to $\mathbf{\Gamma}_p(t + \Delta t)$ \\
\qquad\quad using $\mathbf{F}_p(\mathbf{x})$; \hfill $\triangleright$ Eq.~(\ref{eq:path_integrator_update})
\State Compute $\mathbf{u}_{\alpha}$ by solving $\textbf{Poisson}(\mathbf{m}_{\alpha})$;
\EndFor{}
\end{algorithmic}
\end{algorithm}
\newpage
\section{Validation}\label{sec:validation}

The proposed particle flow map solver has been validated using various benchmark flows. In all experiments, we utilize the \(4^{\text{th}}\)-order Runge–Kutta (RK4) scheme for advection and a Multigrid Preconditioned Conjugate Gradient (MGPCG) solver for the Poisson projection \cite{mcadams2010parallel}. Viscosity is treated explicitly as an arbitrary force, using the fourth-order Laplacian kernel proposed in \cite{cross2020monotonicity}. 

In the following sections, we compare our method with established Eulerian-Lagrangian approaches, including Vortex-In-Cell (VIC) \cite{rossinelli2008vortex}, Fluid-Implicit-Particle (FLIP) \cite{brackbill1986flip, zhu2005animating}, and Affine Particle-In-Cell (APIC) \cite{jiang2015affine}. Additionally, we evaluate our solver against the third-order Adams–Bashforth scheme combined with the fifth-order WENO \cite{liu1994weighted} / WENO-Z \cite{borges2008improved}. While higher-order advection schemes, such as the fourth-order Adams–Bashforth or advanced convection term calculation techniques like WENO-AO \cite{balsara2016efficient} and Targeted ENO (TENO) \cite{fu2016family, fu2019targeted}, could achieve higher accuracy, they generally incur significantly greater computational costs compared to simpler particle-based methods. 

Although our method is not intended as a replacement for high-order schemes, it is, to the best of our knowledge, the first hybrid Eulerian-Lagrangian approach that achieves convergence rates and accuracy comparable to these schemes without using high-order interpolation kernel like WENO or MLS as used in HOPIC\cite{edwards2012high}. This balance between computational efficiency and the ability to accurately calculate physical properties suggests that the proposed method has strong potential for a variety of applications.

All simulations are performed on a GPU platform using 32-bit floating-point precision and executed on an Nvidia RTX 4090 GPU.

\subsection{2D Taylor-Green Vortex}

\begin{figure}[!htb]
    \centering
    \includegraphics[width=0.99\linewidth]{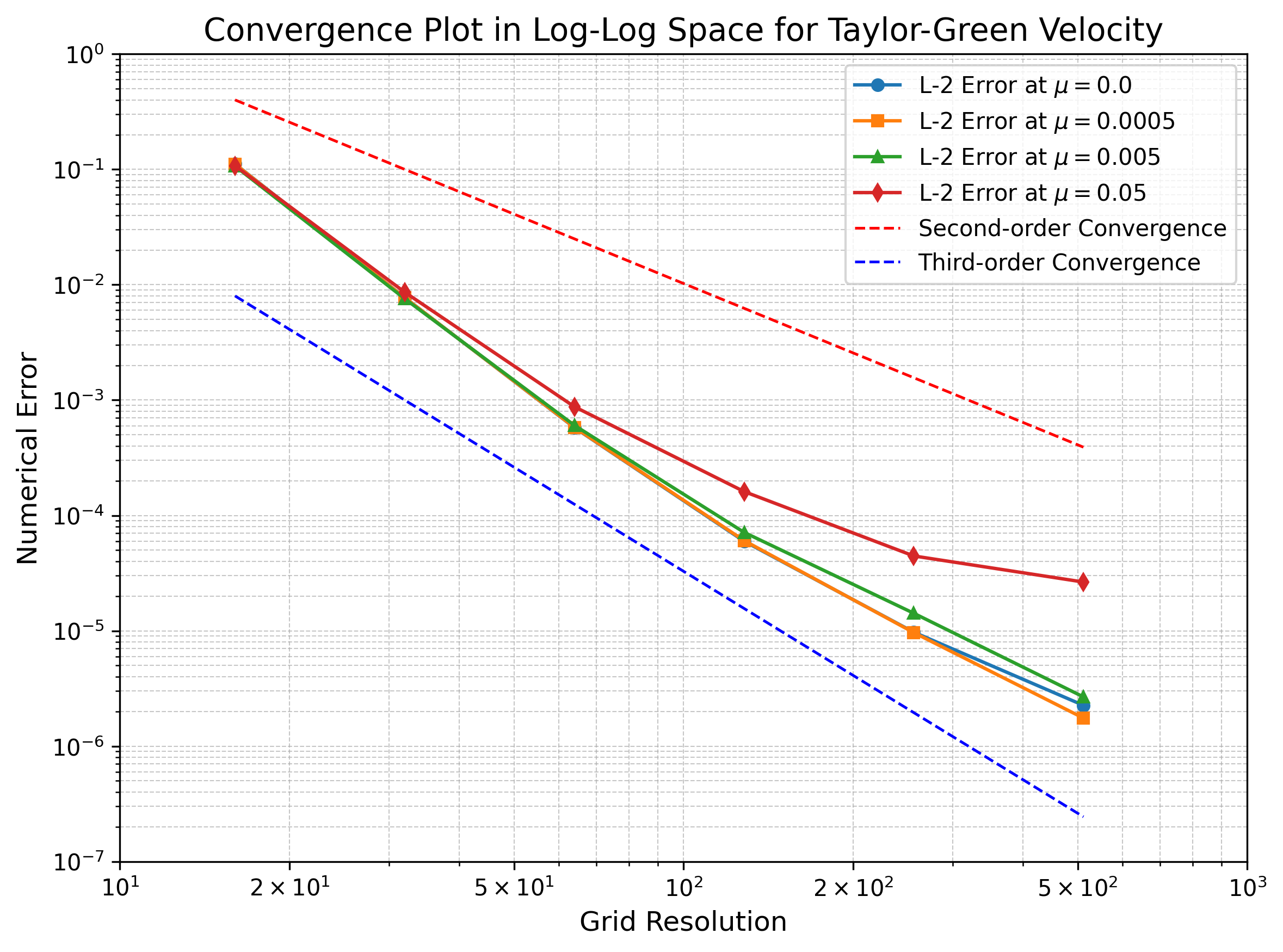}
    \caption{This figure shows the convergence plot of $L_2$ error of our method simulated for the Taylor Green Vortex comparing against the analytical solution. X-axis shows the grid resolution used for computation for the domain size of $2\pi$.}
    \label{fig:2d_taylor_green_convergence_l2}
\end{figure}

\begin{figure}[!htb]
    \centering
    \includegraphics[width=0.99\linewidth]{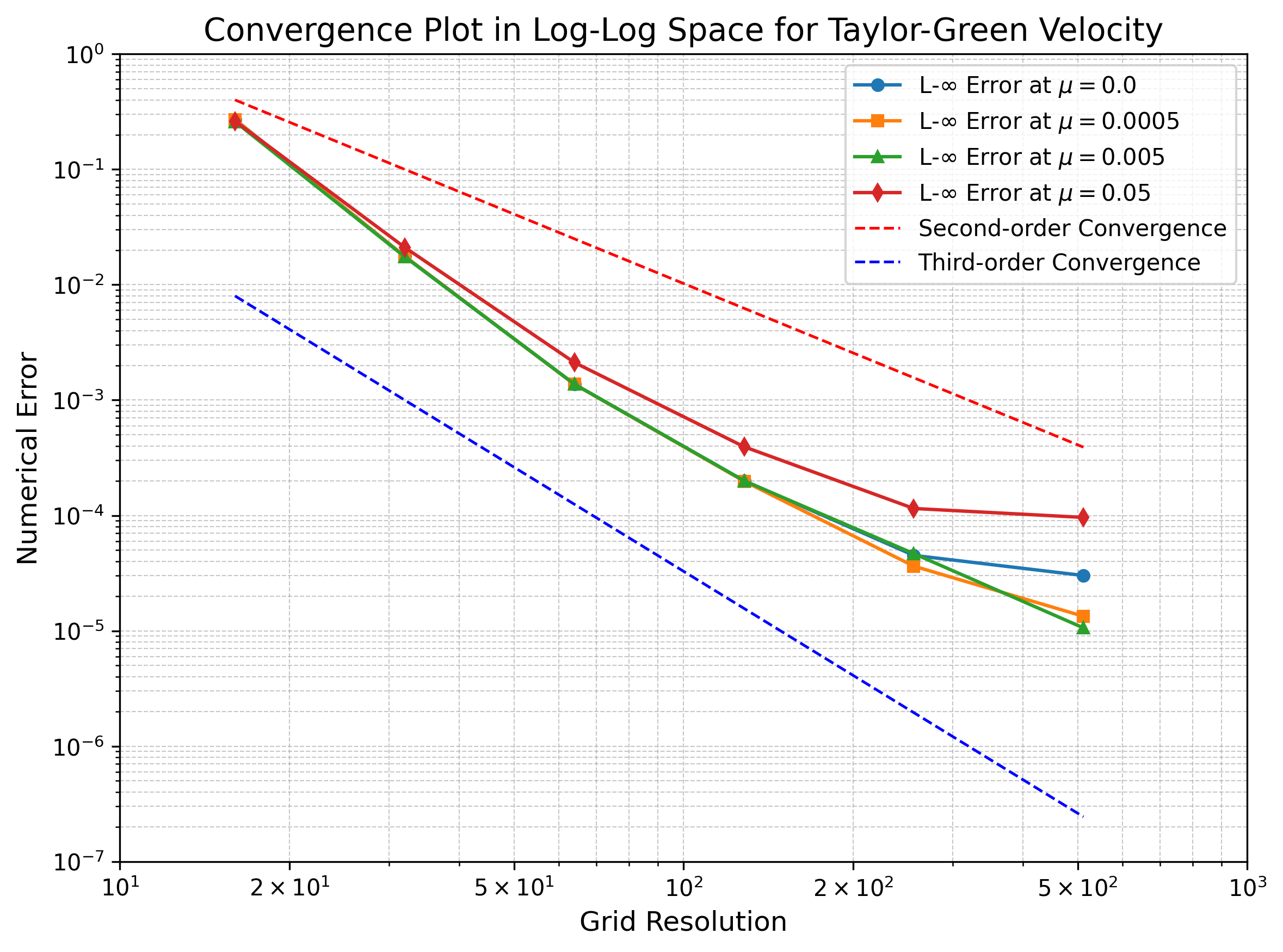}
    \caption{This figure shows the convergence plot of $L_{\infty}$ error of our method simulated for the Taylor Green Vortex comparing against the analytical solution. X-axis shows the grid resolution used for computation for the domain size of $2\pi$.}
    \label{fig:2d_taylor_green_convergence_inf}
\end{figure}

\begin{figure}[!htb]
    \centering
    \includegraphics[width=0.99\linewidth]{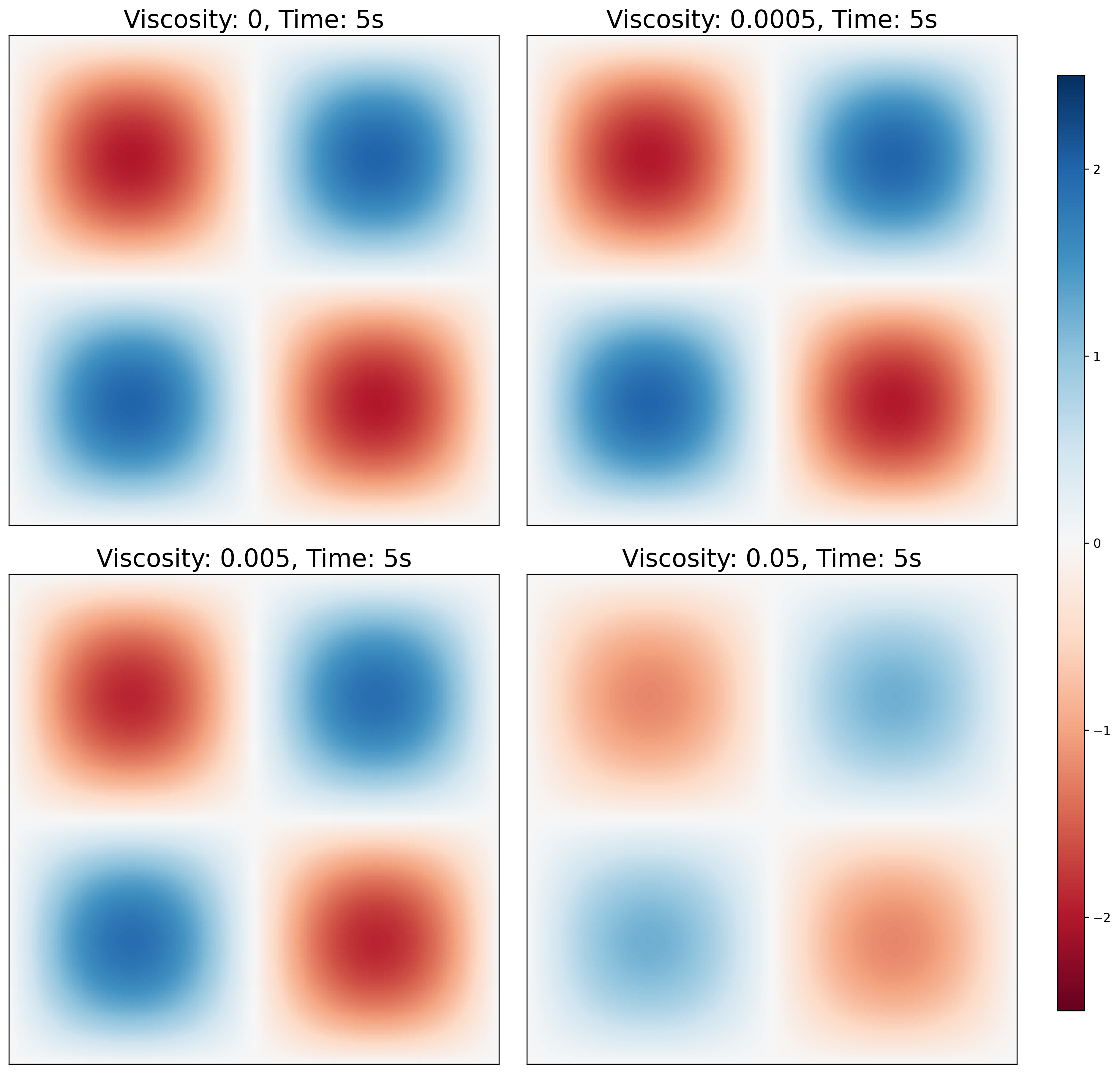}
    \caption{This figure shows the Taylor Green Vortex simulation at time $t=5s$ with different viscosity.}
    \label{fig:2d_taylor_green_vort}
\end{figure}

To evaluate the convergence rate of our method, we compared its results with the analytical solution of the two-dimensional Taylor–Green vortex problem for the Navier–Stokes equations under varying viscosity settings. The velocity field for this benchmark is defined as:  
\[
u(x, y, t) = \sin(x) \cos(y) \, e^{-2\nu t}, \quad 
v(x, y, t) = -\cos(x) \sin(y) \, e^{-2\nu t},
\]
where \(u\) and \(v\) denote the velocity components in the \(x\) and \(y\) directions, respectively; \(\nu\) represents the kinematic viscosity; and \(t\) is the elapsed time. 

Simulations were performed over a domain of \(2\pi \times 2\pi\) for a duration of 5 seconds, with viscosity coefficients of \(0\), \(5 \times 10^{-4}\), \(5 \times 10^{-3}\), and \(5 \times 10^{-2}\) and the results are shown in Fig.~\ref{fig:2d_taylor_green_vort}. The \(L_2\) and \(L_\infty\) errors were computed against the analytical solution to assess accuracy. Periodic boundary conditions were applied throughout, with a CFL number of 1 for all cases except \(\nu = 5 \times 10^{-2}\), where a CFL number of 0.5 was used due to the explicit handling of viscosity. 

Our results demonstrate that the method achieves third-order convergence for low-viscosity flows as shown in Fig.~\ref{fig:2d_taylor_green_convergence_l2} and Fig.~\ref{fig:2d_taylor_green_convergence_inf}. At higher viscosities, the convergence order decreases to between second and third order. This reduction is primarily attributed to the explicit treatment of viscosity and the lack of higher-order accurate discretization for the viscous terms.

\subsubsection{Comparison with other methods}
\begin{figure}[!htb]
    \centering
    \includegraphics[width=0.99\linewidth]{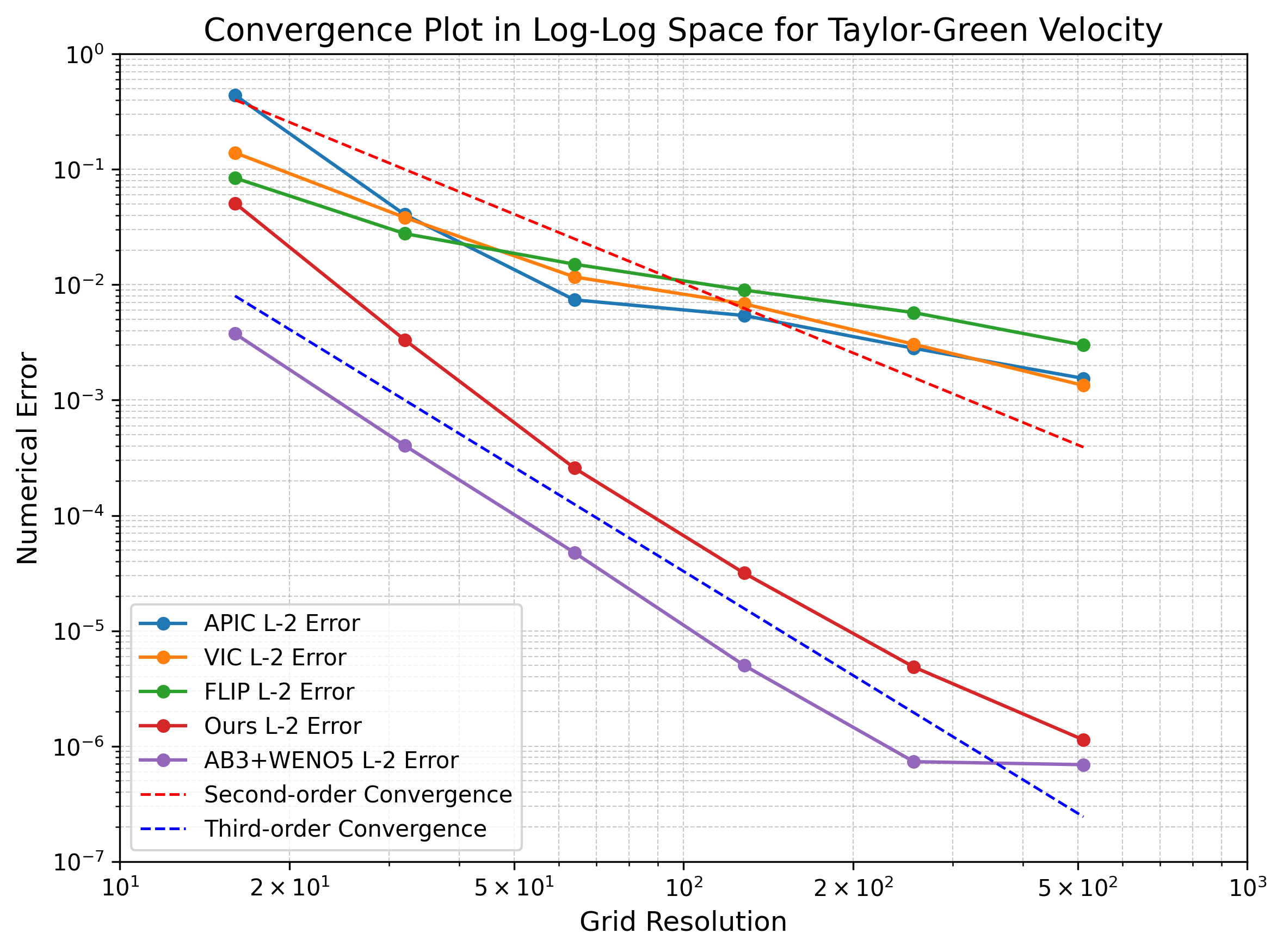}
    \caption{We compare our method with various existing Eulerian Lagrangian methods. Our method shows a third order convergence comparable with AB3+WENO5 while other hybrid methods shows superlinear but below second-order convergence rate.}
    \label{fig:comparison_green}
\end{figure}

Under the same experimental settings, we compare our method with AB3+WENO5, VIC, APIC and FLIP. The results are shown in Fig.~\ref{fig:comparison_green}. 

For the inviscid Taylor-Green vortex simulation, our method shows a higher convergence rate compared to the hybrid methods (VIC, FLIP, and APIC). When compared to the grid-based AB3+WENO5 method, our method achieves a similar convergence order, albeit with a higher absolute error value. It is worth noting that the drop in convergence rate observed at a resolution of 512 for both AB3+WENO5 and our method can likely be attributed to the use of floating-point precision on the GPU.

\subsection{2D Leapfrogging Vortex}
\begin{figure}[!htb]
    \centering
    \includegraphics[width=0.99\linewidth]{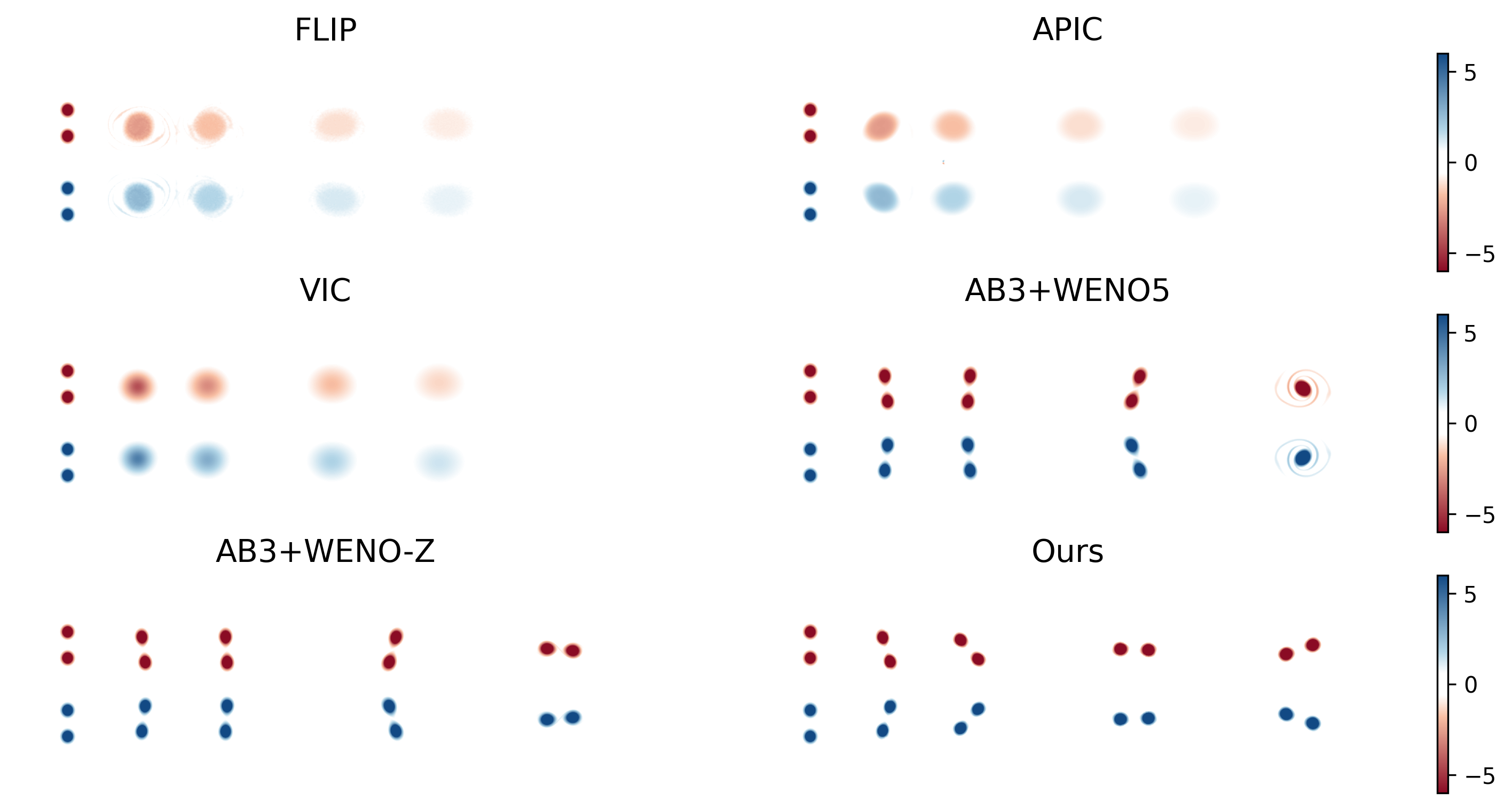}
    \caption{In this figure, we perform a 2D leapfrogging vortex experiment to test the ability of vorticity preservation of a group of methods including FLIP, VIC, APIC, AB3+WENO5, AB3+WENO-Z, and ours. Our method performs slightly better than AB3+WENO5 in the vorticity preservation case and gives a similar result as AB3+WENO-Z. Compared with hybrid methods, our method gives the best result.}
    \label{fig:2d_leapfrog}
\end{figure}
To evaluate the method's ability to preserve vorticity, we conduct the classic 2D leapfrog experiment by initializing four point vortices centered at \(x = 0.25\) and \(y = 0.26, 0.38, 0.62,\) and \(0.74\). Each vortex has the same strength (magnitude) of 0.005, with the upper two vortices being negative and the lower two positive. The velocity fields are computed using a mollified Biot-Savart kernel with support 0.02. 

The computational domain is set to \(256 \times 768\) to ensure that the vortices not interacting with the domain boundaries before merging, allowing sufficient time to test the vorticity preservation capability of the method. Vorticity snapshot at time $t=20s$, $40s$, $70s$, and $120s$ are captures. We compare our approach with the third-order Adams–Bashforth scheme combined with WENO5\cite{liu1994weighted} and WENO-Z \cite{borges2008improved}, as well as the hybrid Eulerian–Lagrangian methods VIC, APIC, and FLIP. 

As shown in Fig.~\ref{fig:2d_leapfrog}, our method demonstrates performance comparable to the AB3+WENO-Z approach and preserves vorticity slightly better due to dissipation in WENO5 as pointed out in \cite{borges2008improved} and yields the best results among the hybrid Eulerian–Lagrangian methods tested in terms of vorticity preservation.

\subsection{2D Kármán Vortex}
We verify the accuracy of our treatment of arbitrary forces within the flow map framework using impulse variables by benchmarking against the 2D Kármán vortex street flow. Specifically, we compute the drag coefficient (\(C_d\)) and lift coefficient (\(C_l\)) across different Reynolds numbers and compare the results with reference data from previous literature.

\subsubsection{Immersed Boundary Method}

We adopt the modified immersed boundary method described in \cite{silva2003numerical}. Following this approach, four immersed boundary forces exerted by static solid boundaries on the flow are computed: the acceleration force \(\bm{f}_a\), inertial force \(\bm{f}_i\), viscous force \(\bm{f}_v\), and pressure force \(\bm{f}_p\). These force components are defined as:

\begin{equation}
\left\{
    \begin{aligned}
    \bm{f}_a(\bm{x}_k) &= \rho \frac{\partial \bm{V}}{\partial t} (\bm{x}_k)\\
    \bm{f}_i(\bm{x}_k) &= \rho (\bm{V} \cdot \nabla) \bm{V} (\bm{x}_k) \\
    \bm{f}_v(\bm{x}_k) &= -\mu \nabla^2 \bm{V} (\bm{x}_k),\\
    \bm{f}_p(\bm{x}_k) &= \nabla P (\bm{x}_k),
    \end{aligned}
    \right.
\end{equation}

where \(\bm{x}_k\) represents the \(k^{\text{th}}\) Lagrangian point of the immersed structure, and \(\bm{V}\) is the velocity on the Eulerian grid. The derivatives are computed using a second-order Lagrange polynomial approximation, as described in \cite{silva2003numerical}.

Due to the use of a gauge transformation to solve the Navier–Stokes equations with impulse variables, the pressure \(P\) is not directly available. The Poisson projection mapping \(\bm{m}\) to \(\bm{V}\) involves a gauge defined as \(\psi\), making \(P\) inaccessible. To address this, at \(t+1\), we perform a single Runge–Kutta 4 (RK4) semi-Lagrangian advection step to compute an intermediate velocity field \(\bm{V}^*\) from \(\bm{V}^n\), which is derived from \(\bm{m}^n\). We then solve a separate Poisson projection to map \(\bm{V}^*\) to \(\bm{V}\), allowing us to extract \(P\) for the calculation of \(\bm{f}_p\). Additionally, to align with the viscosity computations used throughout this paper, we employ the fourth-order Laplacian kernel proposed in \cite{cross2020monotonicity} instead of the original viscosity calculation in \cite{silva2003numerical}.

\subsubsection{Experiment Details}

The flow past a stationary circular cylinder was simulated within a rectangular domain. The boundary conditions were configured so that the flow entered from the bottom and exited from the top. A circular cylinder with diameter \(d = 1/30\) was positioned at coordinates \(x = 0.5\) and \(y = 1.1\). The domain had a length of \(2\) and a width of \(1\). These dimensions were chosen to minimize boundary effects on the flow. Neumann boundary conditions were imposed on the lateral boundaries, and a uniform velocity profile \(U_\infty\) was prescribed at the domain entrance. The Reynolds number is defined as:

\[
Re = \frac{U_\infty \rho d}{\mu},
\]

and simulations were performed for \(\text{Re} = 10, 20, 40, 50, 60, 80, 100, 150, \text{and } 300\) using a fixed resolution of \(\Delta x = 1/256\). 

The number of Lagrangian points was determined based on the criterion \(\Delta s / \Delta x \leq 0.9\). 

Once the velocity and pressure fields were computed, the drag and lift coefficients were calculated using the force field. Following \cite{silva2003numerical}, the drag coefficient \(C_d\) is defined as:

\begin{equation}
    C_d = \frac{F_d}{\frac{1}{2} \rho U_\infty^2 d},
\end{equation}

where \(F_d\) is the drag force, calculated using the force field stored on the grid as:

\begin{equation}
    F_d = - \int_\Omega F_x \, dx,
\end{equation}

with \(F_x\) being the \(x\)-component of the Eulerian forces, and \(\Omega\) the computational domain. Similarly, the lift coefficient \(C_l\) is given by:

\begin{equation}
    C_l = \frac{F_l}{\frac{1}{2} \rho U_\infty^2 d},
\end{equation}

where the lift force \(F_l\) is computed as:

\begin{equation}
    F_l = - \int_\Omega F_y \, dy,
\end{equation}

with \(F_y\) representing the \(y\)-component of the Eulerian forces.

To avoid the instability associated with explicit force addition in the immersed boundary method, we follow the recommendation in \cite{silva2003numerical} by initializing the time step with \(\Delta t = 1 \times 10^{-5}\) and not exceeding \(\Delta t = 1 \times 10^{-3}\). The simulations were run for at least 200 dimensionless time units, as suggested in \cite{silva2003numerical}.

\begin{figure}[!htb]
    \centering
    \includegraphics[width=0.99\linewidth]{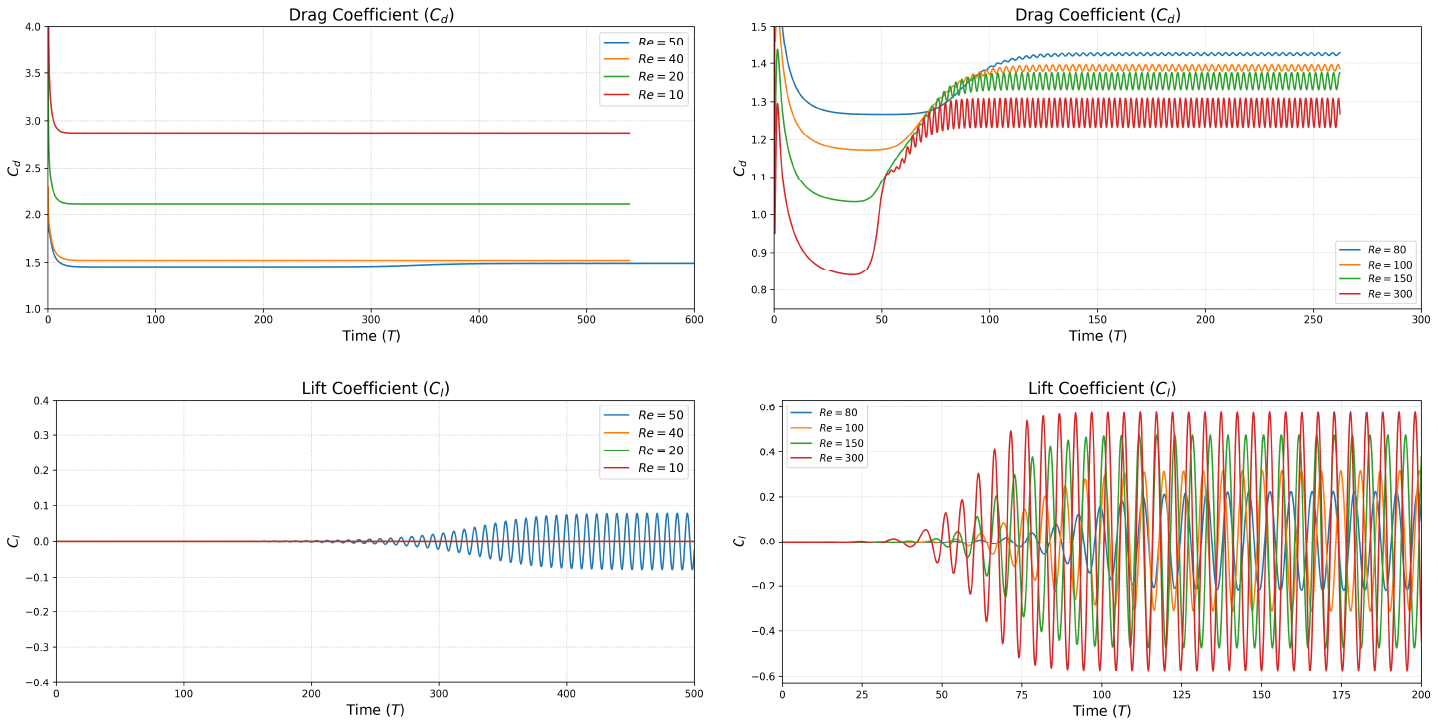}
    \caption{We show the $C_d$ and $C_l$ profile with time evolution. Our result highly resembles the curve given by \cite{silva2003numerical}.}
    \label{fig:2d_karman_vortex}
\end{figure}

In Table~\ref{tab:karman_comparison}, we present the drag coefficient \(C_d\) computed using our method and compare it against reference data from previous studies, showing strong agreement. Figure~\ref{fig:2d_karman_vortex} illustrates the temporal profiles of the drag and lift coefficients, which align closely with those reported in \cite{silva2003numerical}.

\begin{table}[!htb]
    \centering
    \scriptsize
    \caption{\textbf{Comparison of $C_d$ at various Reynolds numbers.}}
    \begin{tabular}{|c|c|c|c|c|c|c|c|}
        \hline
        Re & Ours & Silva et al. \cite{silva2003numerical} & Park et al. \cite{park1998numerical} & Sucker et al. \cite{sucker1975fluiddynamik} & Dennis et al. \cite{dennis1970numerical} & Ye et al. \cite{ye1999accurate} & Tritton \cite{triton1959experiments} \\
        \hline
        10  & 2.87 & 2.81 & 2.78 & 2.67 & 2.05 & 2.03 & 2.22 \\
        20  & 2.11 & 2.04 & 2.01 & 2.08 & 2.05 & 2.03 & 2.22 \\
        40  & 1.52 & 1.54 & 1.51 & 1.73 & 1.52 & 1.52 & 1.48 \\
        47  & -    & 1.46 & -    & -    & -    & -    & -    \\
        50  & 1.49 & 1.46 & -    & 1.65 & -    & -    & -    \\
        80  & 1.42 & 1.40 & 1.35 & 1.51 & -    & -    & 1.29 \\
        100 & 1.39 & 1.39 & 1.33 & 1.45 & -    & -    & -    \\
        150 & 1.35 & 1.37 & -    & 1.36 & -    & -    & -    \\
        300 & 1.27 & 1.27 & 1.37 & 1.22 & -    & 1.38 & -    \\
        \hline
    \end{tabular}
    \label{tab:karman_comparison}
\end{table}

\subsection{2D Cavity Flow}
\begin{figure}[!htb]
    \centering
    \includegraphics[width=0.99\textwidth]{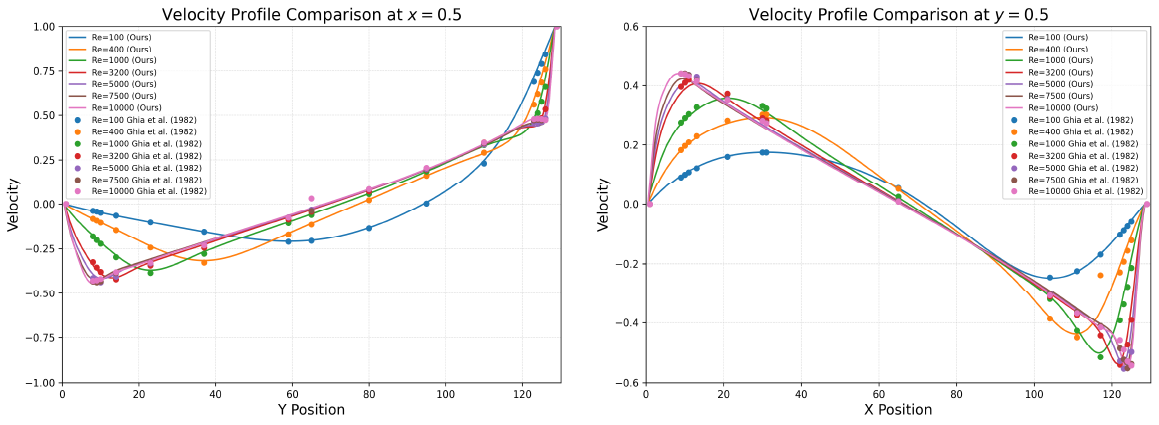}
    \caption{We show the velocity profile at $x=0.5$ and $y=0.5$ for different Reynolds number comparing with the referenced solution from \cite{ghia1982high}.}
    \label{fig:cavity}
\end{figure}

\begin{figure}[!htb]
    \centering
    \includegraphics[width=0.99\textwidth]{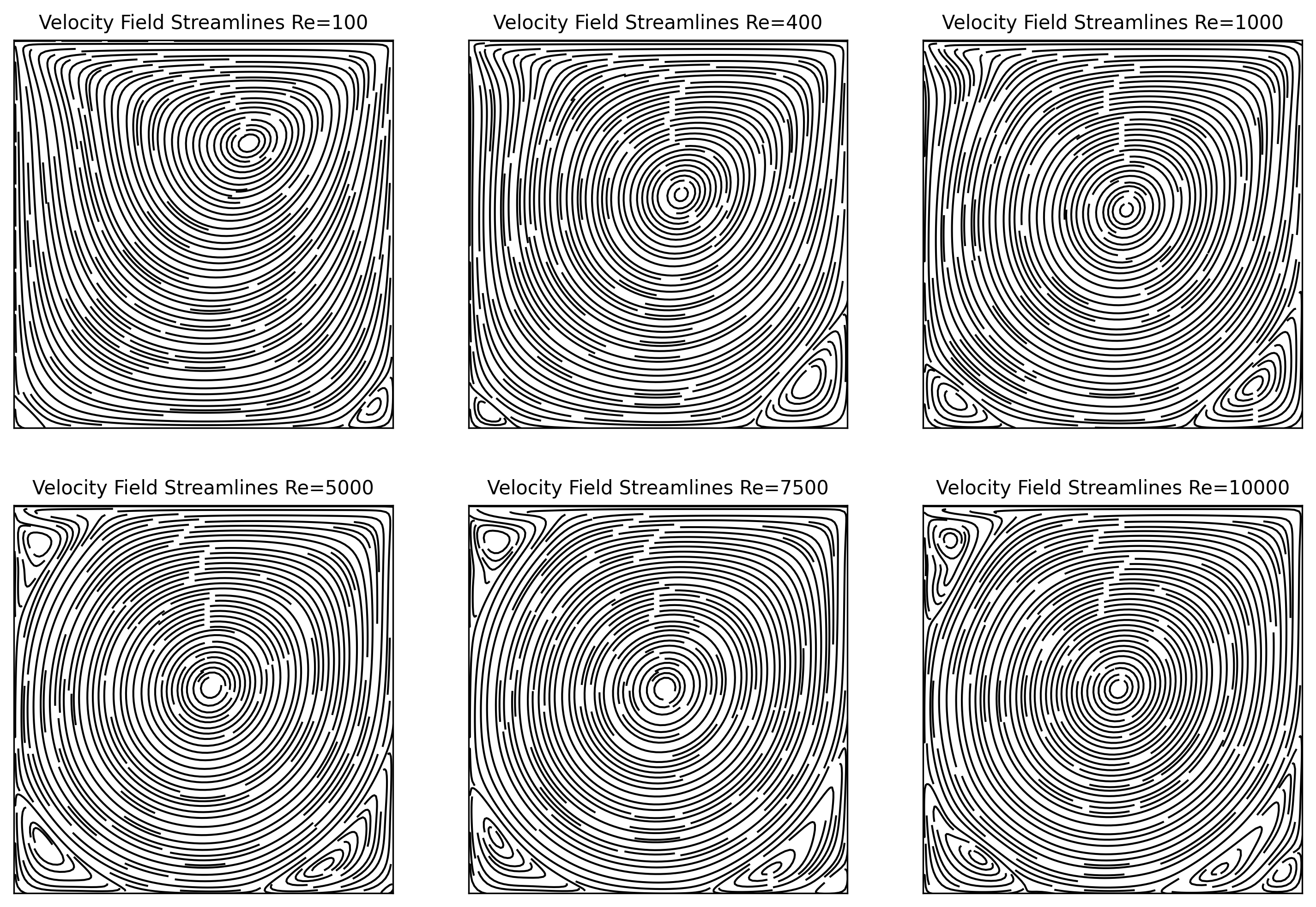}
    \caption{We show the streamline plot of the velocity field after converging. The secondary vortex at a high Reynolds number aligns with the result given in \cite{ghia1982high}.}
    \label{fig:cavity_stream}
\end{figure}
We applied our method to simulate the classical 2D lid-driven cavity flow benchmark case at various Reynolds numbers and compared the results to reference solutions provided by \cite{ghia1982high}. In this setup, the tangential velocity of the top lid is set to 1, while no-slip boundary conditions are imposed on the remaining three walls.

Figure~\ref{fig:cavity} presents the velocity profiles and flow structures obtained using our method for Reynolds numbers of 100, 400, 1000, 3200, 5000, 7500, and 10000. The results show good agreement with the reference data, even at higher Reynolds numbers where flow structures become more complex. Additionally, the streamline plots in Fig.~\ref{fig:cavity_stream} capture the formation and placement of secondary vortices, consistent with the positions reported in \cite{ghia1982high}.

For further validation, the numerical results are provided in Table~\ref{table:cavity_vertical_line} and Table~\ref{table:cavity_horizontal_line}. These comparisons highlight the robustness and accuracy of our method across a wide range of flow conditions.

\subsubsection{Ablation on viscosity calculation}
\begin{figure}[!htb]
    \centering
    \includegraphics[width=0.99\textwidth]{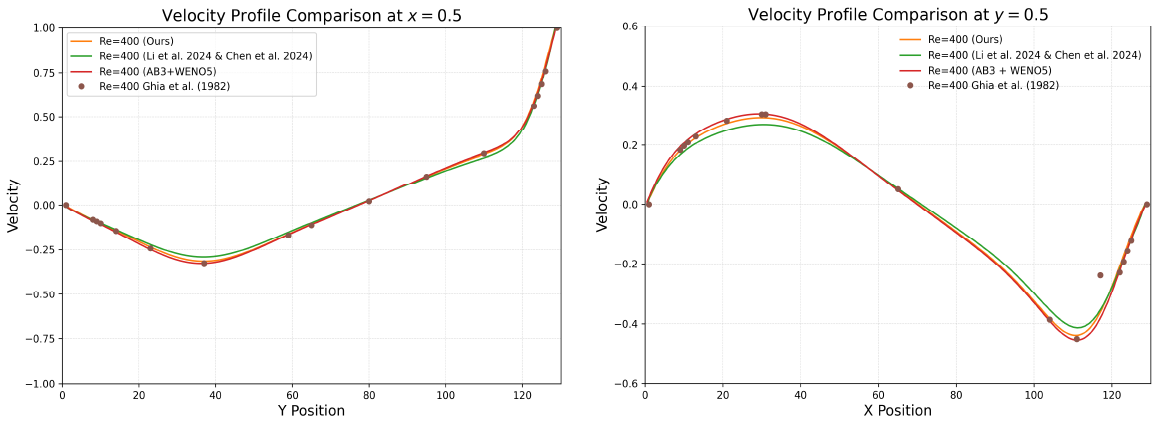}
    \caption{By comparing the velocity profile under considerable viscous force, our result better aligns with the result computed from AB3+WENO5 and the referenced solution comparing with the velocity profile computed using method by \cite{chen2024sfi} and \cite{li2024ink}.}
    \label{fig:comparison_cavity}
\end{figure}
We compare our method with the method from the graphics community for handling external forces on flow maps proposed by Li et al. and Chen et al. \cite{chen2024sfi, li2024ink}.  The results, presented in Fig.~\ref{fig:comparison_cavity}, indicate that our method achieves comparable accuracy to AB3+WENO5 when evaluated against the reference solution provided by \cite{ghia1982high}. The methods proposed by Li et al. and Chen et al. \cite{chen2024sfi, li2024ink} show less accurate results in this comparison. This discrepancy appears to arise from their approach to pressure buffer calculations and their calculation of the viscosity term from the pull-backed velocity field rather than the impulse field, as described in \cite{cortez1995impulse}. This approach introduces numerical errors that impact their accuracy.

% \begin{table}[h]
%     \centering
%     \caption{\textbf{Results for $u$-Velocity along Vertical Line through Geometric Center of Cavity}}
%     \label{cavity_vertical_line}
%     \scriptsize % Make the table smaller
%     \begin{tabular}{|c|c|c|c|c|c|c|c|}
%         \hline
%         $y$ & Re = 100 & Re = 400 & Re = 1000 & Re = 3200 & Re = 5000 & Re = 7500 & Re = 10000 \\ \hline
%         1.0000 & - & - & - & - & 1.00000 & 1.00000 & - \\
%         0.9766 & - & - & - & - & 0.45781 & 0.44226 & - \\
%         0.9688 & - & - & - & - & 0.43045 & 0.43751 & - \\
%         0.9609 & - & - & - & - & 0.42734 & 0.44153 & - \\
%         0.9531 & - & - & - & - & 0.42846 & 0.44200 & - \\
%         0.8516 & - & - & - & - & 0.31402 & 0.31499 & - \\
%         0.7344 & - & - & - & - & 0.18148 & 0.18403 & - \\
%         0.6172 & - & - & - & - & 0.07170 & 0.07470 & - \\
%         0.5000 & - & - & - & - & -0.02674 & -0.02386 & - \\
%         0.4531 & - & - & - & - & -0.06502 & -0.06219 & - \\
%         0.2813 & - & - & - & - & -0.20793 & -0.20454 & - \\
%         0.1719 & - & - & - & - & -0.30359 & -0.29938 & - \\
%         0.1016 & - & - & - & - & -0.38147 & -0.36391 & - \\
%         0.0703 & - & - & - & - & -0.40373 & -0.40893 & - \\
%         0.0625 & - & - & - & - & -0.39352 & -0.41401 & - \\
%         0.0547 & - & - & - & - & -0.37432 & -0.40949 & - \\
%         0.0000 & - & - & - & - & 0.00000 & 0.00000 & - \\ \hline
%     \end{tabular}
% \end{table}

\begin{table}[!htb]
    \centering
    \caption{\textbf{Results for $u$-Velocity along Vertical Line through Geometric Center of Cavity}}
    \label{table:cavity_vertical_line}
    \scriptsize % Make the table smaller
    \begin{tabular}{|c|c|c|c|c|c|c|c|}
        \hline
        $y$ & Re = 100 & Re = 400 & Re = 1000 & Re = 3200 & Re = 5000 & Re = 7500 & Re = 10000 \\ \hline
        1.0000 & 1.00000 & 1.00000 & 1.00000 & 1.00000 & 1.00000 & 1.00000 & 1.00000 \\
        0.9766 & 0.84380 & 0.76533 & 0.66057 & 0.51590 & 0.46951 & 0.45202 & 0.47163 \\
        0.9688 & 0.79211 & 0.68991 & 0.57327 & 0.45958 & 0.44194 & 0.44782 & 0.47899 \\
        0.9609 & 0.74140 & 0.62167 & 0.50669 & 0.43890 & 0.43948 & 0.45277 & 0.48161 \\
        0.9531 & 0.69203 & 0.56149 & 0.45902 & 0.43372 & 0.44105 & 0.45353 & 0.47684 \\
        0.8516 & 0.23666 & 0.28290 & 0.32249 & 0.33212 & 0.32648 & 0.32599 & 0.33278 \\
        0.7344 & 0.00376 & 0.15798 & 0.18176 & 0.19249 & 0.19196 & 0.19312 & 0.19521 \\
        0.6172 & -0.13666 & 0.01970 & 0.05469 & 0.07385 & 0.07660 & 0.07876 & 0.07909 \\
        0.5000 & -0.20449 & -0.11504 & -0.06001 & -0.03472 & -0.02933 & -0.02652 & -0.02607 \\
        0.4531 & -0.20896 & -0.17090 & -0.10435 & -0.07689 & -0.07501 & -0.06747 & -0.06700 \\
        0.2813 & -0.15425 & -0.31721 & -0.27163 & -0.23056 & -0.22009 & -0.21587 & -0.21840 \\
        0.1719 & -0.09993 & -0.23077 & -0.36840 & -0.32916 & -0.31565 & -0.31091 & -0.31836 \\
        0.1016 & -0.06342 & -0.13773 & -0.27824 & -0.41045 & -0.39256 & -0.37508 & -0.38358 \\
        0.0703 & -0.04595 & -0.09734 & -0.20559 & -0.37754 & -0.41180 & -0.41928 & -0.43565 \\
        0.0625 & -0.04138 & -0.08727 & -0.18646 & -0.35432 & -0.39981 & -0.42152 & -0.44421 \\
        0.0547 & -0.03670 & -0.07716 & -0.16696 & -0.32615 & -0.37855 & -0.41310 & -0.44057 \\
        0.0000 & 0.00000 & 0.00000 & 0.00000 & 0.00000 & 0.00000 & 0.00000 & 0.00000 \\ \hline
    \end{tabular}
\end{table}

\begin{table}[!htb]
    \centering
    \caption{\textbf{Results for $v$-Velocity along Horizontal Line through Geometric Center of Cavity}}
    \label{table:cavity_horizontal_line}
    \scriptsize % Make the table smaller
    \begin{tabular}{|c|c|c|c|c|c|c|c|c|}
        \hline
        $x$ & Re = 100 & Re = 400 & Re = 1000 & Re = 3200 & Re = 5000 & Re = 7500 & Re = 10000 \\ \hline
        1.0000 & 0.00000 & 0.00000 & 0.00000 & 0.00000 & 0.00000 & 0.00000 & 0.00000 \\
        0.9688 & -0.06171 & -0.12060 & -0.21374 & -0.39628 & -0.47016 & -0.51823 & -0.54495 \\
        0.9609 & -0.07710 & -0.15557 & -0.27477 & -0.47710 & -0.52745 & -0.53984 & -0.55522 \\
        0.9531 & -0.09229 & -0.19104 & -0.33287 & -0.52422 & -0.53561 & -0.51363 & -0.51515 \\
        0.9453 & -0.10719 & -0.22630 & -0.38511 & -0.53833 & -0.51324 & -0.47527 & -0.47152 \\
        0.9063 & -0.17433 & -0.37545 & -0.50189 & -0.42500 & -0.40102 & -0.39939 & -0.41051 \\
        0.8594 & -0.22920 & -0.43861 & -0.41205 & -0.35886 & -0.35085 & -0.34730 & -0.35573 \\
        0.8047 & -0.24803 & -0.37581 & -0.30674 & -0.29748 & -0.28891 & -0.28647 & -0.29302 \\
        0.5000 & 0.05660 & 0.05295 & 0.02560 & 0.01440 & 0.01201 & 0.01001 & 0.00747 \\
        0.2344 & 0.17575 & 0.29122 & 0.31221 & 0.27408 & 0.26269 & 0.25723 & 0.25822 \\
        0.2266 & 0.17553 & 0.29137 & 0.32009 & 0.28203 & 0.27037 & 0.26478 & 0.26608 \\
        0.1563 & 0.16106 & 0.27078 & 0.35578 & 0.35954 & 0.34126 & 0.33431 & 0.33810 \\
        0.0938 & 0.12329 & 0.22104 & 0.31166 & 0.40539 & 0.40949 & 0.40474 & 0.40637 \\
        0.0781 & 0.10899 & 0.20145 & 0.28982 & 0.39266 & 0.41158 & 0.41992 & 0.42676 \\
        0.0703 & 0.10100 & 0.18994 & 0.27692 & 0.38090 & 0.40597 & 0.42235 & 0.42676 \\
        0.0625 & 0.09241 & 0.17708 & 0.26224 & 0.36570 & 0.39515 & 0.41914 & 0.43938 \\
        0.0000 & 0.00000 & 0.00000 & 0.00000 & 0.00000 & 0.00000 & 0.00000 & 0.00000 \\ \hline
    \end{tabular}
\end{table}

\subsection{3D Arnold-Beltrami-Childress Test}
\begin{figure}[!htb]
    \centering
    \includegraphics[width=0.99\linewidth]{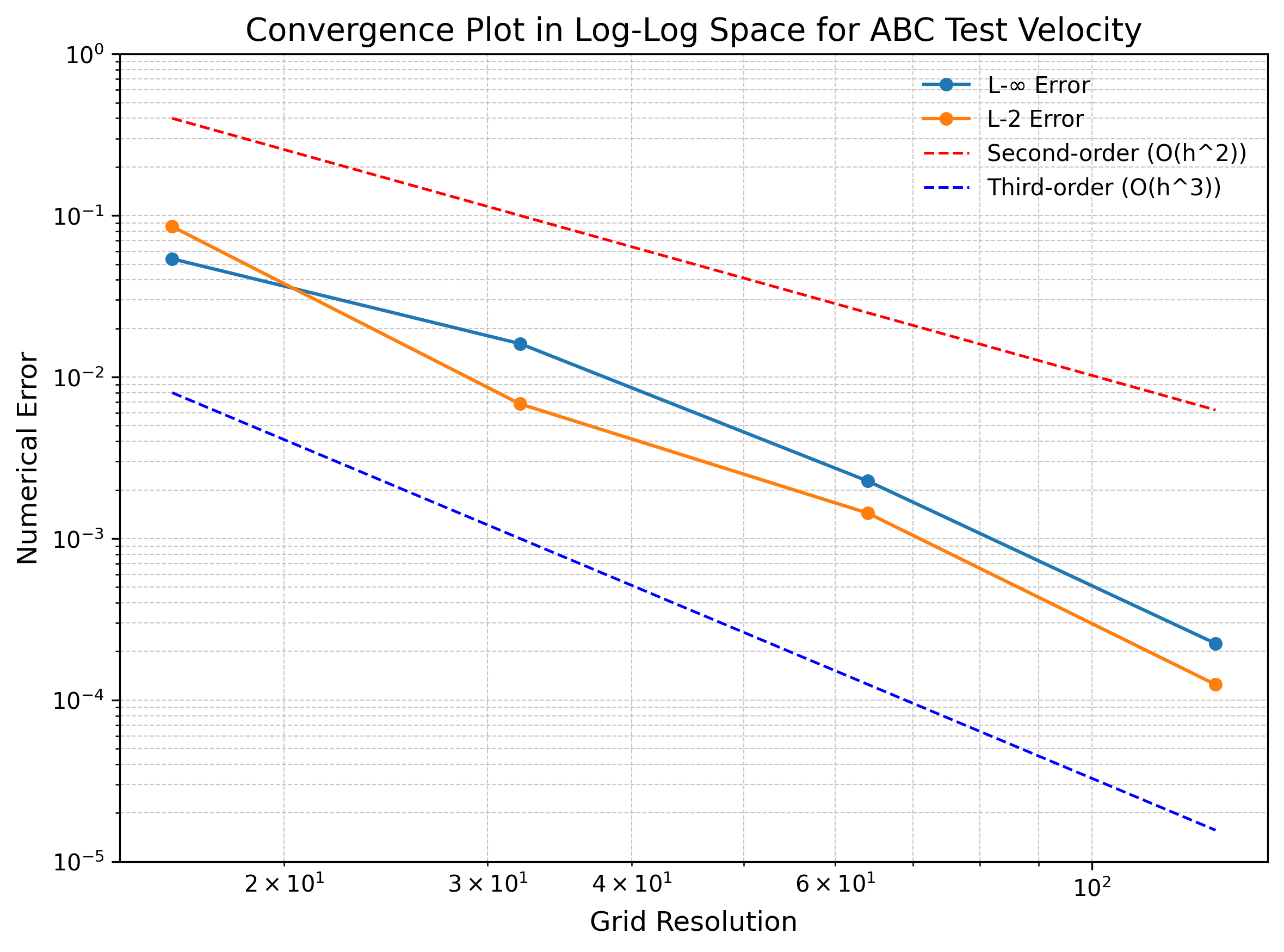}
    \caption{We show the convergence rate of our method under a 3D ABC test. Third order convergence can be observed for both $L_2$ and $L_{\infty}$ errors.}
    \label{fig:3d_abc_convergence_inf}
\end{figure}

\begin{figure}[!htb]
    \centering
    \includegraphics[width=0.99\linewidth]{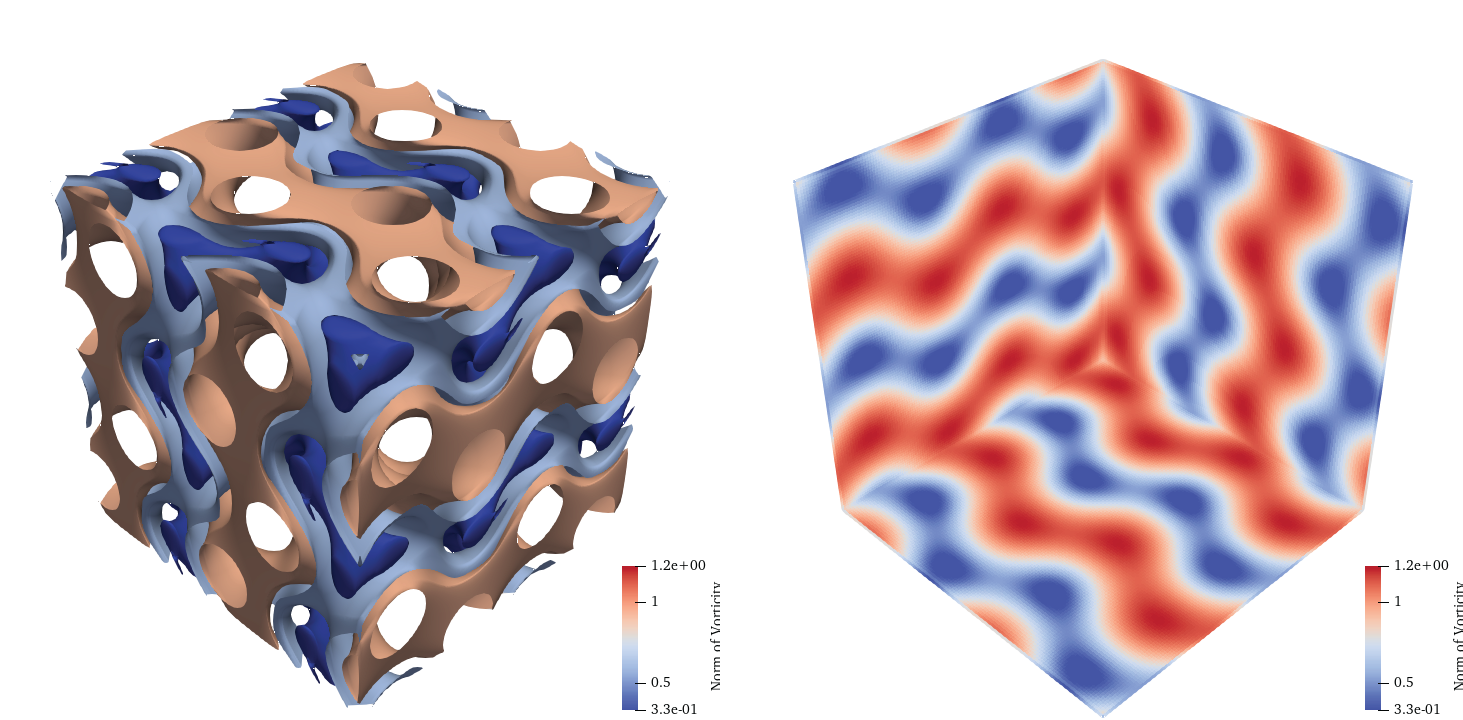}
    \caption{This figure shows the vorticity isosurface and vorticity slices of 3D ABC velocity field after simulated for 5 seconds.}
    \label{fig:3d_abc_vort}
\end{figure}
We conducted an Arnold-Beltrami-Childress (ABC) flow test with the initial velocity field defined as:

\[
\mathbf{u}_0(x, y, z) = \frac{1}{2}
\begin{pmatrix}
\cos(y) + \sin(z) \\
\cos(z) + \sin(x) \\
\cos(x) + \sin(y)
\end{pmatrix}.
\]

The simulation was performed in a periodic cubic domain of size \([ -2\pi, 2\pi ]^3\), with a grid spacing \(\Delta x = \frac{4 \pi}{N}\), where \(N\) denotes the grid resolution. A CFL number of 0.5 was used, and the simulation was run for \(t = 5\,\text{s}\). The \(L_2\) and \(L_\infty\) errors were computed relative to the reference solution at the initial time frame.
We show our simulation result in Fig.~\ref{fig:3d_abc_vort} and our results demonstrate third-order convergence as shown Fig.~\ref{fig:3d_abc_convergence_inf}.

\subsection{3D Leapfrogging Vortex}

\begin{figure}
    \centering
    \includegraphics[width=0.99\linewidth]{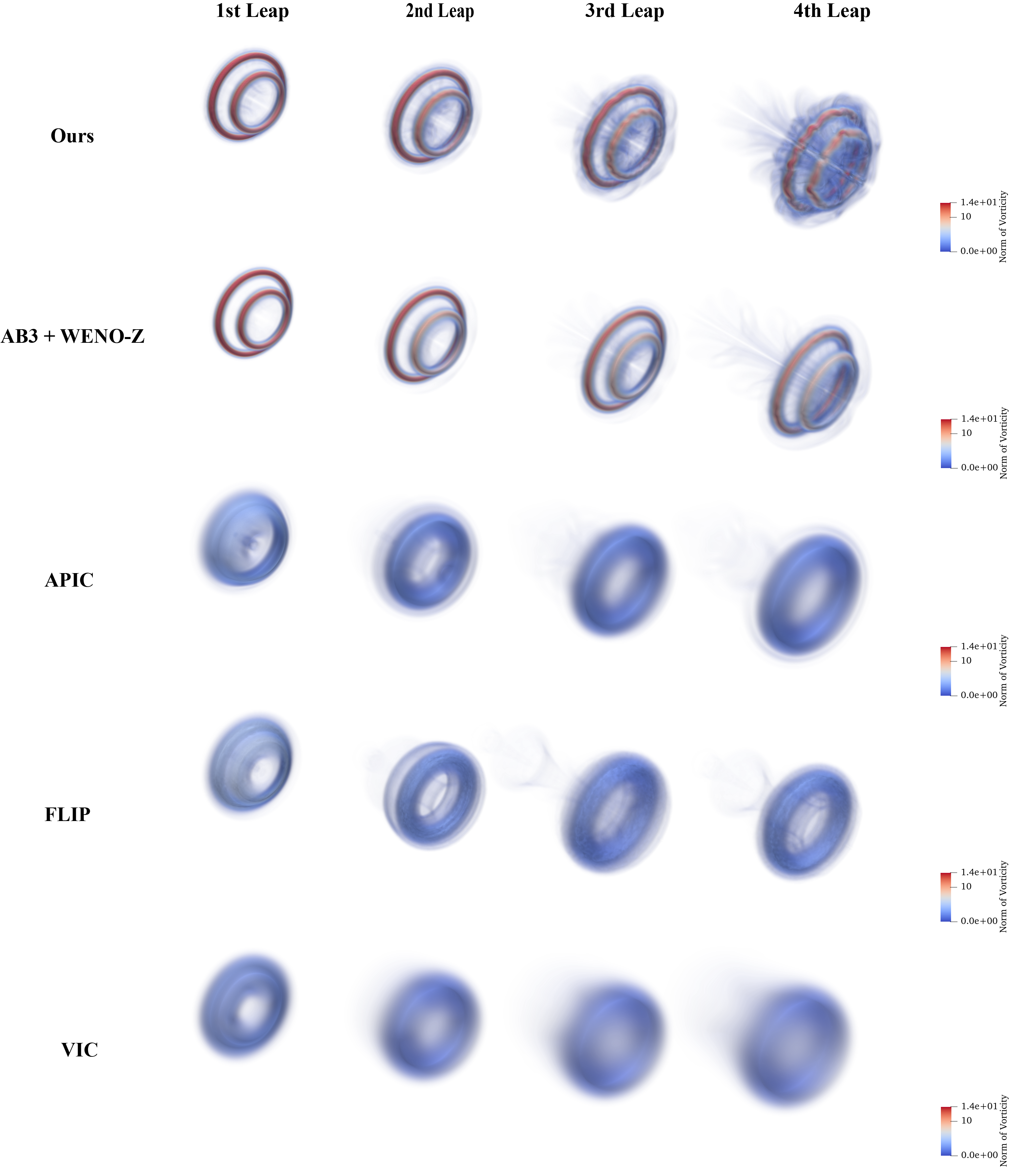}
    \caption{We compare our method with various hybrid methods to show the vorticity preservation strength. Our method gives vortex rings remaining splited after the fourth leap which is comparable with the result obtained through AB3+WENO5.}
    \label{fig:3d_leapfrog}
\end{figure}
To further investigate the ability of our method on vorticity preservation under the 3D setting, we perform the 3D leapfrogging test. We initialize two parallel vortex rings with their centers positioned at \(x = 0.16\) and \(x = 0.29125\), while the \(y\)- and \(z\)-coordinates are fixed at 0.5. The major radius of the vortex rings is set to 0.21, and the minor radius, representing the mollification support of the vortices, is 0.0168. 

When comparing our method to hybrid particle methods such as APIC, VIC, and FLIP, we observe that it preserves the separation between the two vortex rings even after the fourth leap. This performance is comparable to the result acquired through using the AB3+WENO-Z method.

\section{Result}\label{sec:result}
In this section, we perform some numerical experiments to further illustrate the capacity of our method in various simulation settings in terms of initial condition, vorticity preservation, and solid boundary handling.

\subsection{3D Taylor-Green Vortex}
\begin{figure}[!htb]
    \centering
    \includegraphics[width=0.99\linewidth]{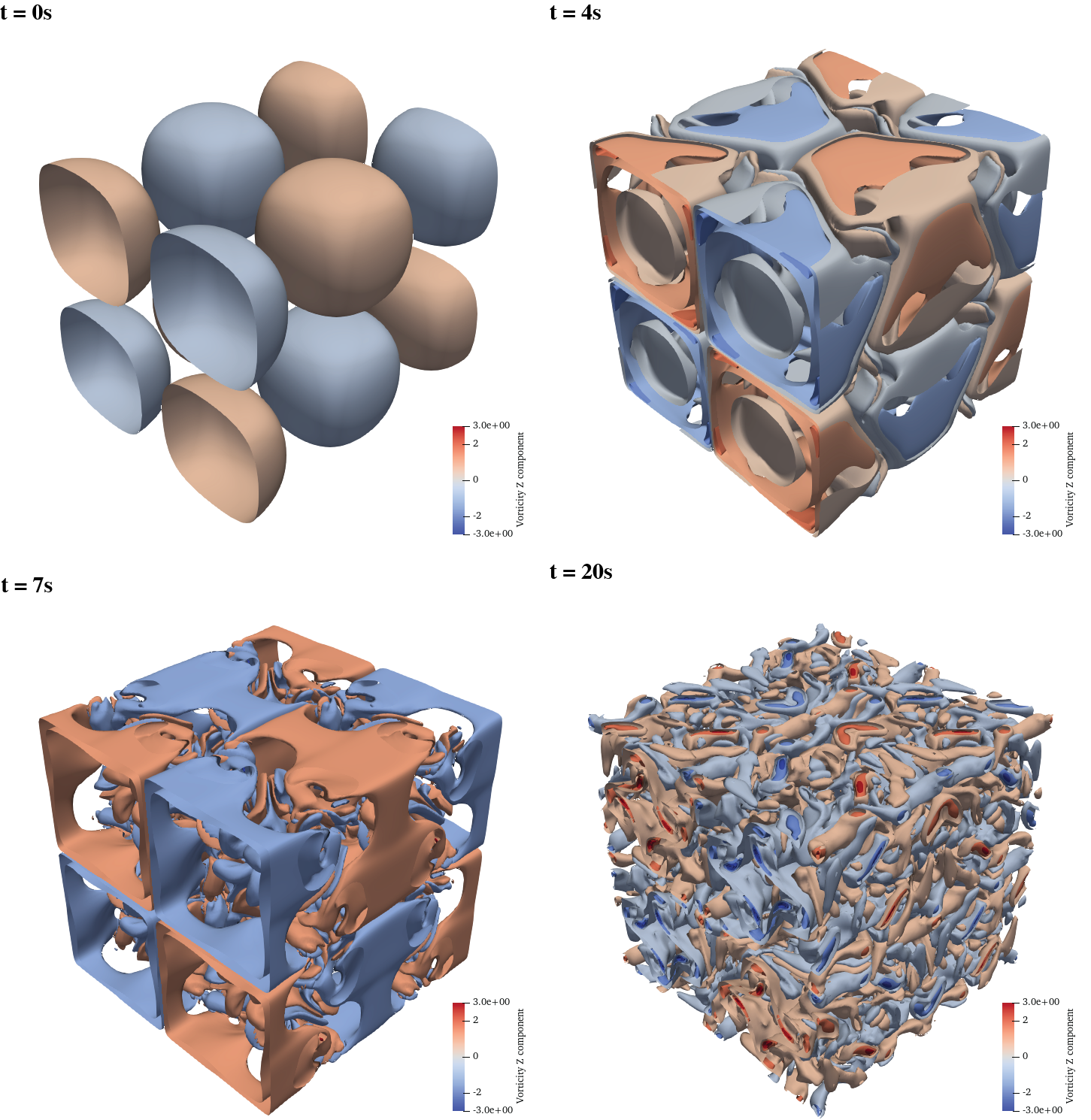}
    \caption{This figure shows the development of a 3D Taylor-Green Vortex transitioning from the initial setting to the final turbulant state.}
    \label{fig:3d_taylor_green}
\end{figure}
In this section, we present the simulation of the three-dimensional Taylor-Green vortex. The velocity field is defined as:

\begin{equation}
\left\{
    \begin{aligned}
        &u = V_0 \sin\left(\frac{x}{L}\right) \cos\left(\frac{y}{L}\right) \cos\left(\frac{z}{L}\right),\\
        &v = -V_0 \cos\left(\frac{x}{L}\right) \sin\left(\frac{y}{L}\right) \cos\left(\frac{z}{L}\right),\\
        &w = 0,
    \end{aligned}
    \right.
\end{equation}

where \(V_0\) is the initial velocity amplitude, and \(L\) denotes the characteristic length scale. The Reynolds number is computed as \(\text{Re} = \frac{V_0 L}{\mu}\), and for this study, we select \(\text{Re} = 1600\).

Figure 5 illustrates the evolution of the flow field, with snapshots taken at \(t = 0\), \(t = 4\), \(t = 7\), and \(t = 20\) seconds. At \(t = 0\), the velocity field shows smooth and symmetric vortex structures typical of the initial Taylor-Green vortex configuration. Over time, vortex interactions result in a progressive breakdown of this initial arrangement into increasingly complex and chaotic structures. By \(t = 20\), the flow exhibits small-scale features characteristic of a turbulent regime.

These results suggest that the numerical method effectively captures the transition of the Taylor-Green vortex from a laminar configuration to a more turbulent state. The findings indicate that the approach is suitable for simulating the evolution of complex three-dimensional flows.

\subsection{20 Vortex Ring Collision}

\begin{figure}[!htb]
    \centering
    \includegraphics[width=0.99\linewidth]{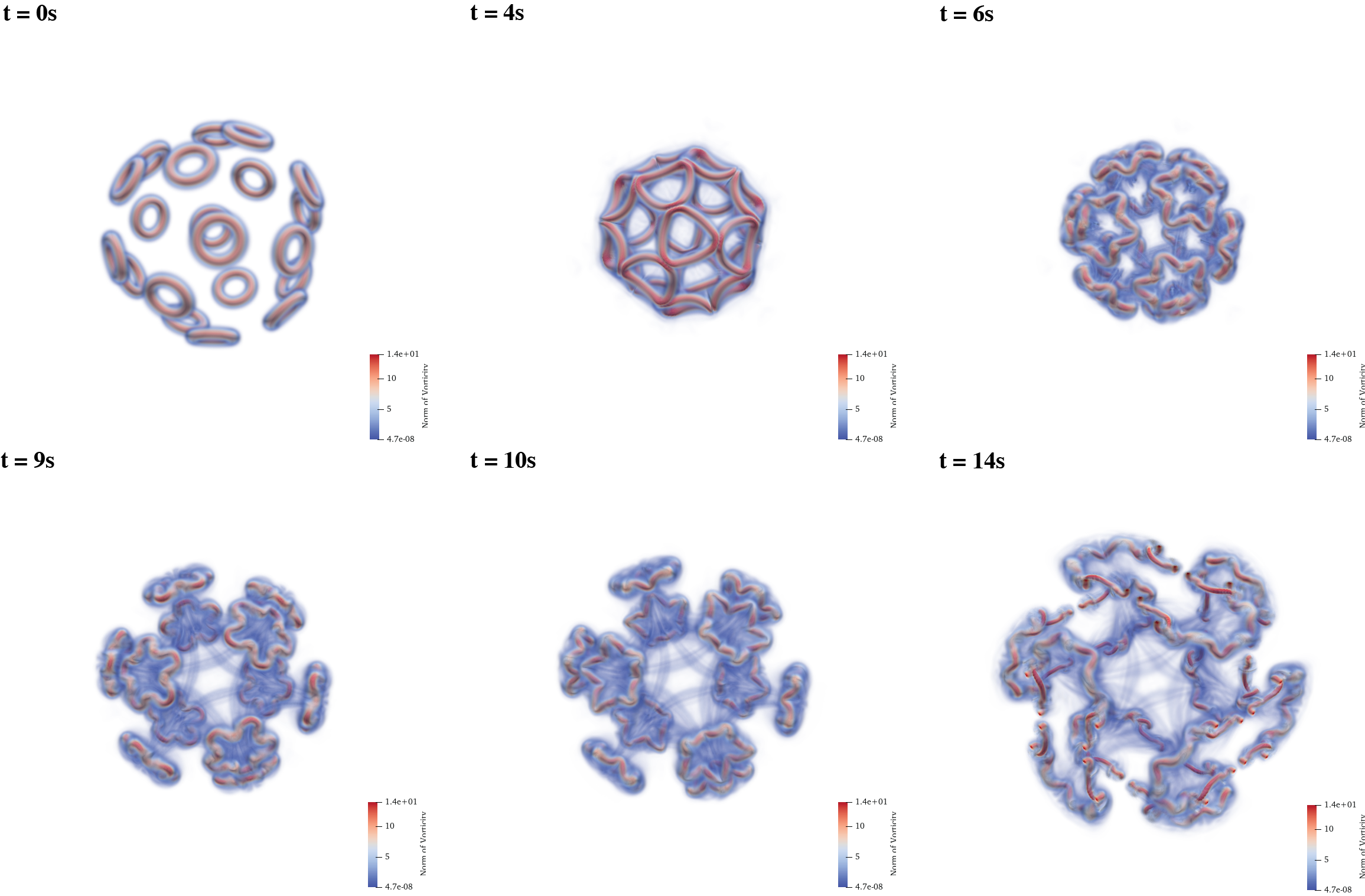}
    \caption{This shows our method simulating 20 vortex rings colliding each other forming semetric vortex structures until colliding with the walls.}
    \label{fig:3d_20_vort}
\end{figure}
In this experiment, vortex rings were initialized to form an icosahedral configuration, centered at \([0.5, 0.5, 0.5]\) within the computational domain \([0, 1]^3\). The The simulation was performed using a CFL number of 0.5 throughout. To ensure numerical stability and accuracy, the flow map reinitialization frequency was set to 15 time steps. The results are presented for \(t = 0, 4, 6, 9, 10, \text{and } 14\) seconds.

The simulation captures the symmetric evolution of the vortex structures as they progress through reconnection and splitting phenomena. These dynamics highlight the method's capability to accurately resolve complex vortex interactions while preserving symmetry and capturing fine-scale details in the flow. The observed reconnections and splits are consistent with expected vortex dynamics, further validating the robustness and accuracy of our approach in simulating three-dimensional turbulent flows.

\subsection{3D Wind Tunnel}

\begin{figure}[!htb]
    \centering
    \includegraphics[width=0.99\linewidth]{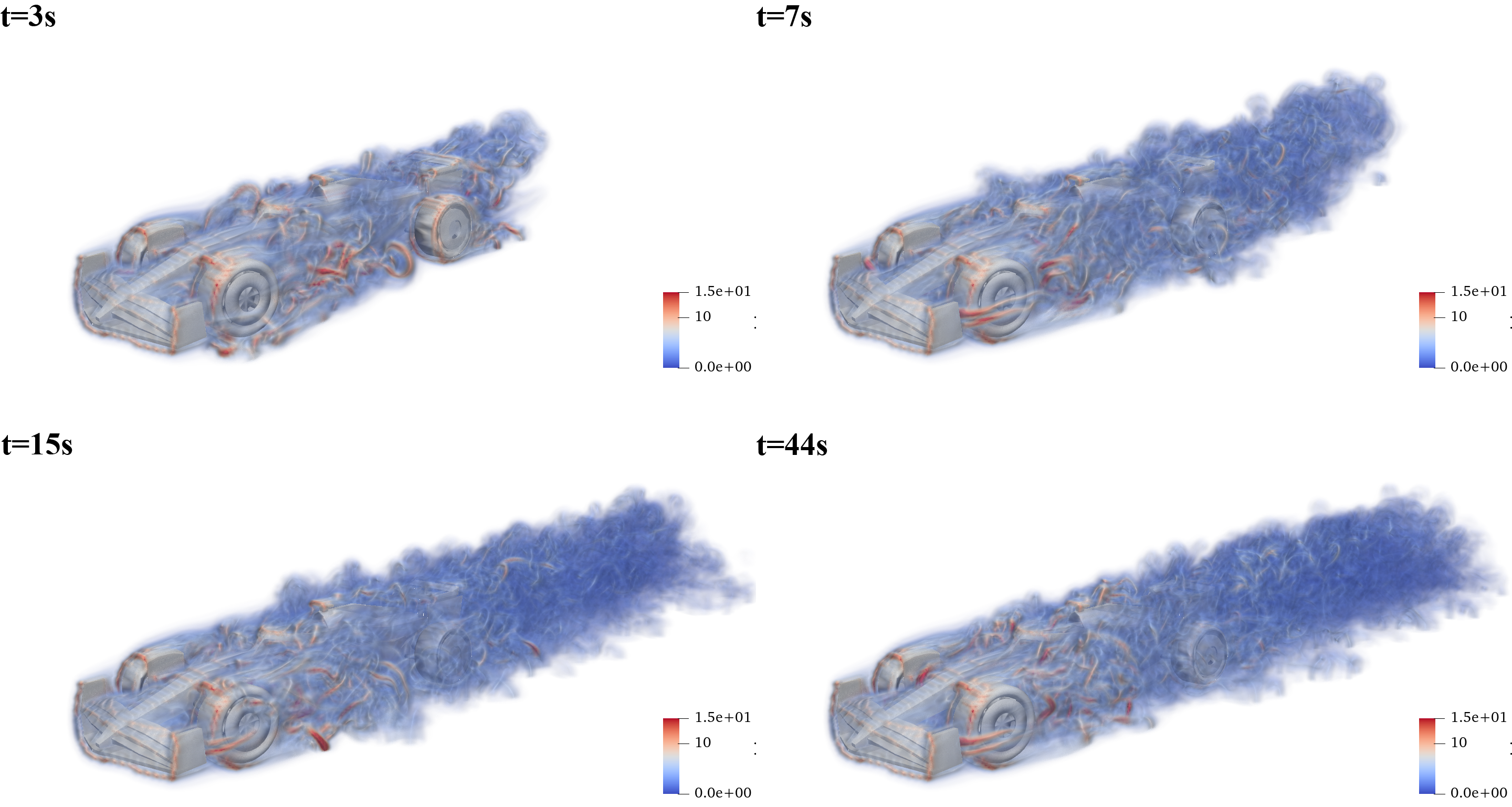}
    \caption{We show our method applied to real-world simulation for a wind tunnel test where an F-1 race car is placed against an incoming flow with constant velocity. Turbulent vortex structures can be observed.}
    \label{fig:3d_car_vort}
\end{figure}

\begin{figure}[!htb]
    \centering
    \includegraphics[width=0.99\linewidth]{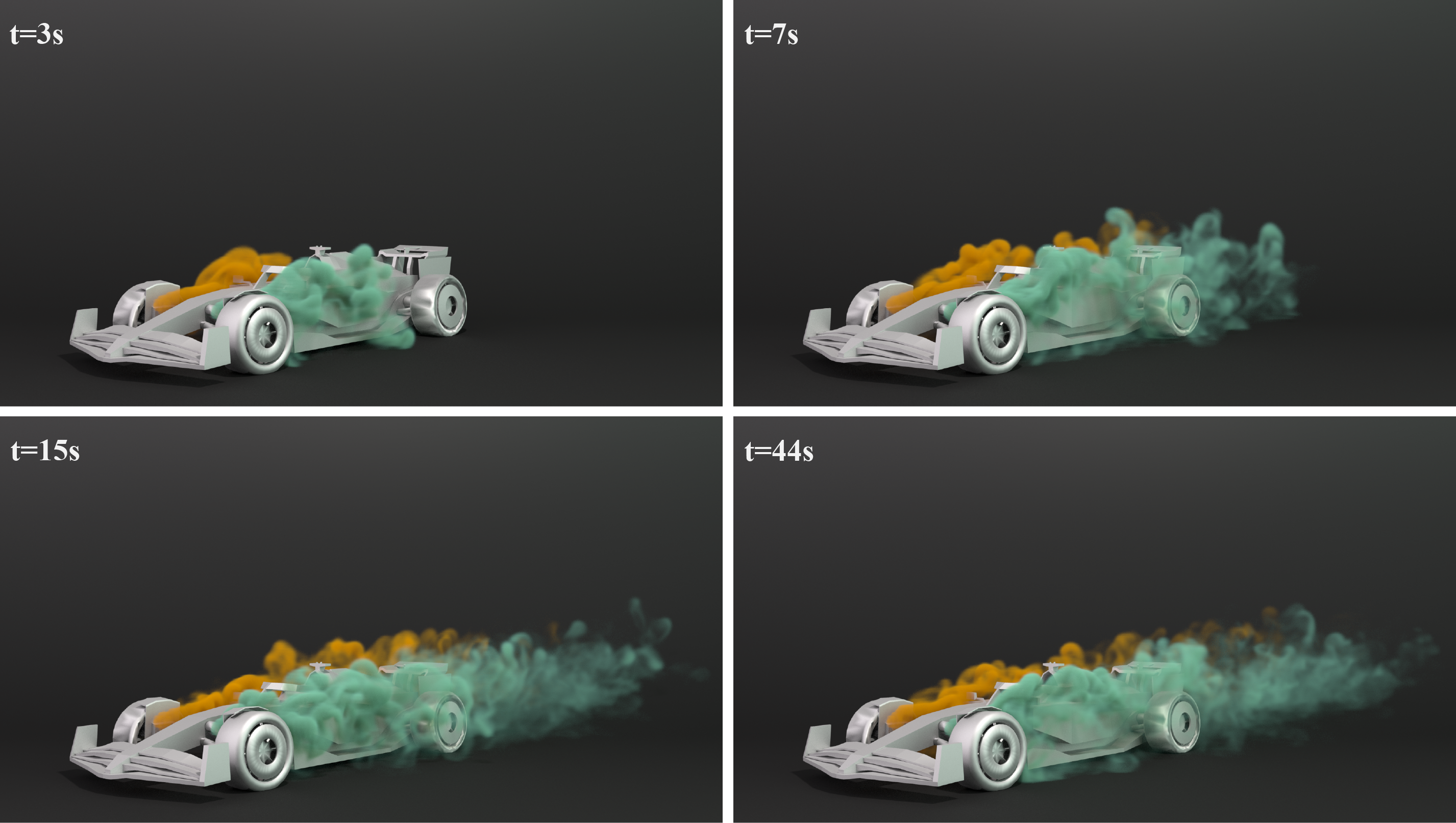}
    \caption{We use smoke to visualize the turbulent flow created on and at the back of the F-1 race car.}
    \label{fig:3d_car_smoke}
\end{figure}

In our final experiment, we evaluate the performance of our fluid solver in a wind tunnel simulation. The computational domain is defined as \(384 \times 128 \times 128\), with wall boundary conditions applied along the \(y\)- and \(z\)-directions, an inlet boundary condition at the front, and an outlet boundary condition at the back. The incoming flow velocity is set to 0.2. The test object is an F-1 race car model positioned at \([0.25, 0.5, 0.5]\) within the domain. The mesh is discretized onto the computational grid, with solid boundary conditions enforced throughout the simulation.

Figure~\ref{fig:3d_car_vort} shows the evolution of the vorticity profiles over time at \(t = 3\), \(t = 7\), \(t = 15\), and \(t = 44\) seconds. At early stages (\(t = 3\) and \(t = 7\)), coherent vortices begin to form around the car, especially near the rear wing and wheels, reflecting the generation of wake structures. As time progresses (\(t = 15\) and \(t = 44\)), these vortices develop into more complex patterns, with the wake extending further downstream. By \(t = 44\), the simulation captures fine-scale turbulent structures and vortex interactions in the wake region, indicating the onset of fully developed turbulence.

The visualized vorticity patterns demonstrate the ability of the solver to capture fine details in turbulent flows and maintain stability in the presence of complex geometry and boundary conditions. These results suggest the solver's suitability for simulating real-world aerodynamic scenarios involving intricate wake dynamics. 

At the end, we show a visualization of the turbulent flow visualized and captured through two colored smoke plume as shown in Fig.~\ref{fig:3d_car_smoke}

\section{Conclusion}

In this work, we presented a particle flow map-based solver for the incompressible Navier-Stokes equations, utilizing an impulse-variable formulation. The method integrates the flow map approach with a time-split framework to handle advection, viscous forces, and external forces. Through numerical experiments, we explored the solver’s performance on a variety of benchmark cases, including the Taylor-Green vortex, Lid-driven cavity flow, Karman vortex street formation and ABC test.

The results suggest that the proposed method provides reasonable accuracy and stability across the tested scenarios. Comparisons with established methods such as VIC, FLIP, APIC, and AB3+WENO5 indicate that our solver achieves competitive performance in vorticity preservation, energy conservation, and capturing complex flow features. Notably, the method demonstrated the ability to handle challenging simulations, such as vortex interactions and turbulent transitions, while maintaining computational efficiency.

While the results are promising, there are limitations that warrant further investigation. For instance, the method’s performance under higher Reynolds numbers or in multiphase flows remains to be thoroughly evaluated. Additionally, integrating higher-order advection schemes or adaptive meshing could potentially enhance accuracy, although these enhancements may come with increased computational costs.

In conclusion, the particle flow map-based solver shows potential for addressing a range of fluid dynamics problems, offering a balance between computational efficiency and accuracy. Future work will aim to address the identified limitations and explore more complex applications.

\section{Acknowledgment}
Georgia Tech authors acknowledge NSF IIS
\#2433322, ECCS \#2318814, CAREER \#2433307, IIS \#2106733, OISE
\#2433313, CNS \#2450401, and CNS \#1919647 for funding support. We credit the
Houdini education license for video animations and renderings.

\newpage
%\appendix

%\section{Sample Appendix Section}

 \bibliographystyle{elsarticle-num} 
 \bibliography{cas-refs,flowmap-refs}
% \bibliography{}




%\input{appendix}

% \end{thebibliography}
\end{document}